\documentclass[pre%
,twocolumn%
,superscriptaddress%
,floats
,aps%
]{revtex4}
\usepackage{amsmath}%

\begin{document}
\title{Numerical Results for Ground States of Spin Glasses on Bethe Lattices} 
\date{\today}
\author{Stefan Boettcher}  
\email{www.physics.emory.edu/faculty/boettcher}
\affiliation{Physics Department, Emory University, Atlanta, Georgia
30322, USA}  

\begin{abstract} 
The average ground state energy and entropy for $\pm J$ spin glasses
on Bethe lattices of connectivities $k+1=3\ldots,26$ at $T=0$ are
approximated numerically. To obtain sufficient accuracy for large
system sizes (up to $n=2048$), the Extremal Optimization heuristic is
employed which provides high-quality results not only for the ground
state energies per spin $e_{k+1}$ but also for their entropies
$s_{k+1}$. The results show considerable quantitative differences
between lattices of even and odd connectivities. The results for the
ground state energies compare very well with recent one-step replica
symmetry breaking calculations. These energies can be scaled for all
{\it even} connectivities $k+1$ to within a fraction of a percent onto
a simple functional form,
$e_{k+1}=E_{SK}\sqrt{k+1}-(2E_{SK}+\sqrt{2})/\sqrt{k+1}$, where
$E_{SK}=-0.7633$ is the ground state energy for the broken replica
symmetry in the Sherrington-Kirkpatrick model.  But this form is in
conflict with perturbative calculations at large $k+1$, which do not
distinguish between even and odd connectivities. We find non-zero
entropies $s_{k+1}$ at small connectivities. While $s_{k+1}$ seems to
vanish asymptotically with $1/(k+1)$ for even connectivities, it is
indistinguishable from zero already for odd $k+1\geq9$.  \hfil\break
PACS number(s): 75.10.Nr, 
                02.60.Pn, 
                89.75.-k, 
                05.10.-a 

\end{abstract} 
\maketitle

\section{Introduction}
\label{introduction}
In this paper we study the ground state ($T=0$) properties of $\pm J$
spin glasses on $k+1$-Bethe lattices \cite{MP2}. The Bethe lattices in
this case are $r$-regular graphs\cite{Bollobas} with $r=k+1$. These
are randomly connected graphs consisting of $n$ vertices in which each
vertex has a fixed connectivity of $k+1$. This constraint contrasts
with ``random graphs'' \cite{ER,Bollobas} in which pairs of vertices
are randomly connected, leading to a Poissonian distribution of
connectivities around a mean of $\left< c\right>$; these graphs will
be studied numerically elsewhere \cite{eo_rg}. We explore the
large-$n$ regime of low-connectivity graphs, $k+1=3,\ldots,26$, which
are of great theoretical interest as finite-connected, mean-field
models for low-dimensional lattice spin glasses \cite{VB,MPV}. A great
number of studies have focused on various aspects of this conceptually
simple model to hone the complex mathematical techniques required to
treat disordered systems \cite{MP1,MP2,PT,MP,Motti,LG,DG} or
optimization problems \cite{MPZ,Monasson,Zecchina,Franz,WS,Banavar}.
In this paper we will try to provide a independent numerical check on
the validity and accuracy of those techniques.

Our results, in turn, reflect on the flexibility of the extremal
optimization (EO) heuristic \cite{BoPe1,eo_prl} in finding approximate
but high-quality solutions for ground states of spin glasses on an
arbitrary graphical structure in a reasonable computational
time. These are often NP-hard optimization problems which are believed
to require a computational effort that rises faster than any power of
$n$ to obtain provably exact solutions \cite{GJ}.  Thus, exact methods
as of yet are not able to provide results for large-$n$ problems, at
least not with significant statistics \cite{Kobe}, except for some
special cases \cite{PA,Rinaldi,Liers}. Furthermore, there is only a
small number of capable approximate algorithms available for the study
of $T=0$ properties of spin glasses \cite{Pal,Hartmann,HM}, mostly
restricted to $d$-dimensional lattice models, and EO provides a
distinct alternative which will increase the confidence in the
numerical results available.  In previous papers, we have demonstrated
the capabilities of EO in determining near-optimal solutions by
reproducing existing results for $3d$ and $4d$ spin glasses and
obtaining new results for the coloring problem\cite{eo_prl,eo_col} and
the graph partitioning problem \cite{eo_perc,BoPe1,BoPe2}. The results
in this paper show that EO is not only capable of approximating ground
states well but also of sweeping the entire configuration space
efficiently to determine the degeneracy of ground states
\cite{eo_col}. Unlike other methods, EO never ``freezes'' into local
minima and proves to be limited only by the ability to store new
ground states.

We find that our results for the ground state energies are consistent
with the theoretical results of the one-step replica symmetry-broken
(1RSB) solution of $\pm J$ spin glasses on $k+1$-connected Bethe
lattices. Our numerical result for $k+1=3$ below clearly excludes the
replica symmetric (RS) solution and are consistent with the 1RSB
results \cite{MP2}. Beyond that our results suggest subtle differences
between even and odd values of the connectivity $k+1$, with no obvious
way to continue smoothly between them \cite{eo_bethe}.  These
oscillations may doom perturbative calculation for $k+1\to\infty$
\cite{PT,LG}.  While the expectation has been raised that the entropy
per spin of the ground states in this model should be vanishing for
any $k+1$ \cite{PT}, we find that the entropy is finite and decaying
like $1/(k+1)$ for large, even $k+1$. For odd $k+1$ it is non-zero
only for small values and may be vanishing already beyond some finite,
odd connectivity.

In the following we first introduce the Bethe lattices we used in the
numerical calculations. In Sec.~\ref{EOalgo} we briefly describe the
EO algorithm which is amply discussed elsewhere
\cite{eo_prl,CISE,BoPe1}. In Sec.~\ref{tests} we present a few
simulations to reproduce known results to gauge our procedure. In
Sec.~\ref{numerics} we present our numerical results, followed by an
extensive discussion in Sec.~\ref{discussion}. Some conclusions are
presented in Sec.~\ref{conclusion}.

\section{Spin Glasses on Bethe Lattices}
\label{sg}
Disordered spin systems on random graphs have been investigated as
mean-field models for low-dimensional spin glasses or optimization
problems, since variables are long-range connected yet have a small
number of neighbors. Particularly simple are Bethe lattices of
connectivity $k+1$ \cite{MP,MP1,MP2}, also called fixed-valence or
$r$-regular random graphs~\cite{Banavar,eo_perc,eo_jam,Bollobas}.
These are graphs consisting of $n$ vertices where each vertex
possesses a fixed number $k+1$ of bonds with randomly selected other
vertices. In comparison to the otherwise more familiar random graphs
studied by Erd\"os and Reny \cite{ER,Bollobas}, Bethe lattices at a
given $n$ and $k$ avoid fluctuations in the connectivities of vertices
and in the total number of bonds.

There are slight variations in the generation of Bethe lattices. For
instance, to add a bond one could choose at random two vertices of
connectivities $<k+1$ to link until all vertices are
$k+1$-connected. Instead, we have used the method described in
Ref.~\cite{Bollobas} to generate these graphs. Here, all the terminals
on the vertices form a list of $n(k+1)$ independent variables. For
each added bond two available terminals are chosen at random to be
linked and removed from the list. Furthermore, for algorithmic
convenience, we reject graphs with possess self loops, bonds that
connect two terminals of the same vertex. Multiple bonds between any
pair of vertices are allowed, otherwise it is too hard to generate
feasible graphs for, say, $n=32$ and $k+1=20$. Since $k+1$ remains
finite for $n\to\infty$, the energy and entropy per spin would only be
effected to $O(1/n)$ by the differences between these choices.

Once a graphical instance is generated, we assign randomly chosen but
fixed couplings $J_{i,j}\in\{-1,+1\}$ to existing bonds between
neighboring vertices $i$ and $j$. Each vertex $i$ is occupied by a
spin variable $x_i\in\{-1,+1\}$. The energy of the system is defined
as the difference in number between violated bonds and satisfied
bonds,
\begin{eqnarray}
H=-\sum_{\{bonds\}} J_{i,j} x_i x_j,
\label{Heq}
\end{eqnarray}
and in this paper we will focus on the energy per spin,
\begin{eqnarray}
e_{k+1}(n)={H\over n},
\label{eeq}
\end{eqnarray}
as a function of $k+1$ in the limit of $n\to\infty$. Each instance can have a large degeneracy $\Omega$ in the configurations exhibiting its ground state energy, and we also sample the average entropy,
\begin{eqnarray}
s_{k+1}(n)=\frac{1}{n}\ln\Omega,
\label{seq}
\end{eqnarray}
for these instances.

\section{$\tau$-EO Algorithm for Bethe Lattices}
\label{EOalgo}
The extremal optimization algorithm, $\tau$-EO, which we employ in this
paper, has been discussed previously in \cite{eo_prl}, and in
\cite{eo_jam,BoPe2} with regard to the setting of its one free parameter,
$\tau$. Here, we merely describe the implementation of $\tau$-EO
without further justification.

To obtain the numerical results in Secs.~\ref{tests}-\ref{numerics},
we used the following implementation of $\tau$-EO: For a given spin
configuration on a graph, assign to each spin $x_i$ a ``fitness''
\begin{eqnarray}
\lambda_i=-\#violated~bonds=-0,-1,-2,\ldots,-(k+1),
\end{eqnarray}
so that
\begin{eqnarray}
e=-{1\over2n}\sum_i\lambda_i
\label{lambdaeq}
\end{eqnarray}
is satisfied. Each spin falls into one of only $k+2$ possible
states. Say, currently there are $n_{k+1}$ spins with the worst
fitness, $\lambda=-(k+1)$, $n_{k}$ with $\lambda=-k$, and so on up to
$n_0$ spins with the best fitness $\lambda=0$. Now draw a ``rank'' $l$
according to the distribution
\begin{eqnarray} 
P(l)={\tau-1\over1-n^{1-\tau}} l^{-\tau}\quad(1\leq l\leq n).
\label{taueq}
\end{eqnarray}
Then, determine $0\leq j\leq(k+1)$ such that
$\sum_{i=j+1}^{k+1}n_i<l\leq\sum_{i=j}^{k+1}n_i$. Finally, select any
one of the $n_j$ spins in state $j$ and reverse its orientation {\em
unconditionally.} As a result, it and its neighboring spins change
their fitness. After all the effected $\lambda$'s and $n$'s are
reevaluated, a new spin is chosen for an update.

This EO implementation updates spins with a ($\tau$-dependent) bias
against poorly adapted spins on behalf of Eq.~(\ref{taueq}). This
process is ``extremal'' in the sense that it focuses on atypical
variables, and it forms the basis of the EO method. The only
adjustable parameter in this algorithm is the power-law exponent
$\tau$. For $\tau=0$, randomly selected spins get forced to update,
resulting in a random walk through the configuration space
which would yield poor results. For $\tau\to\infty$, only spins in the
worst state get updated which quickly traps the update process to a
small region of the configuration space which may be far from a
near-optimal solution. The arguments given in \cite{eo_jam} and a few
experiments indicate that $\tau=1.3$ is a good choice to find ground
states efficiently on Bethe lattices.

The algorithm never converges or ``freezes'' into a particular state
but perpetually explores new near-optimal configurations. It is, of
course, easy to simply store the lowest energy state found so far in a
given run of $\tau$-EO and terminate when desired. Previous experience
with optimizing spin glasses with EO \cite{eo_prl}, and a few
experiments, suggest a typical number of updates of $O(n^3)$ for an
EO-run to obtain saturation in the values found for ground states, at
least up to the system sizes $n\approx 10^3$ obtainable here. Instead
of pushing to attain larger values of $n$, we opt here for obtaining
better statistics by sampling more instances at smaller values of $n$
while spending even more time on each instance than may seem to be
required, in an attempt to ensure accuracy. In particular, our
implementation restarts for each instance at least $r_{\rm max}=4$
times with new random initial spin-assignments, executing
$\approx0.1\,n^3$ updates per run. If a new, lower-than-previous
energy state is encountered in run $r$, we adjust $r_{\rm max}=2+2r$
for that instance so that EO runs at least twice as many restarts as
were necessary to find the lowest state in the first place. Especially
for small $n$, $r_{\rm max}$ hardly ever exceeds 4; for
larger $n$ a few graphs require up to 25 restarts before termination.

Since EO perpetually explores new configurations it is well suited to
explore also the degeneracy of low-energy states. In this case we not
only store the first configuration found with the lowest energy for
that instance. Instead, we consider each configuration with the lowest
energy, retaining new ones and rejecting all others. This procedure is
somewhat inefficient and at best allows system sizes up to $n=256$
beyond which the degeneracy exceeds memory constraints. But it
provides a fast way to also determine the $T=0$ entropy of the ground
states with moderate accuracy. In these runs, we used a similar
approach to the above, except for setting $r_{\rm max}=8+2r$ where $r$
is the latest run in which another new configuration of the lowest
energy was located. Here, for some highly degenerate instances at larger $n$,
$r_{\rm max}$ could reach up into the $100$'s, further limiting
attainable system sizes.

\section{Numerical Test}
\label{tests}
To evaluate the proposed $\tau$-EO algorithm, we have run a series of
test. First, we can defer to some already published results
\cite{eo_prl,eo_jam}. In Ref.~\cite{eo_prl} we have calculate
approximations to the ground state energy for $\pm J$ spin glasses on
a hypercubic lattice for $d=3$ and $d=4$ for systems up to
$n=12^3=1728$ which for each $n$ reproduced previous results obtained with
sophisticated genetic algorithms \cite{Pal,Hartmann}
(although there we used a fixed $r_{\rm max}$). To evaluate the
ability of the algorithm to determine the degeneracy of low-energy
states found, we have reproduced within statistical error the results
of Ref.~\cite{Hartmann_entro} up to $n=6^3$ beyond which EO ran out of time
and memory to sample states completely. (Ref.~\cite{Hartmann_entro} used a
more efficient way to estimate the entropy from sampling only a small
number of states.) And it took EO only a fraction of a second to find
all 60 ground states of a $4^3$ instance that had been exactly
enumerated in Ref.~\cite{Kobe}.

To gauge $\tau$-EO's performance for larger $n$, we have run our
implementation also on two $3d$ lattice instances, $toruspm$3-8-50 and
$toruspm$3-15-50, with $n=8^3=512$ and $n=15^3=3375$, considered in
the 7th DIMACS challenge for semi-definite problems
\cite{DIMACS7}. Bounds \cite{Liers_pc} on the ground-state cost
established for the larger instance are $H_{\rm lower}=-6138.02$ (from
semi-definite programming) and $H_{\rm upper}=-5831$ (from
branch-and-cut). EO found $H_{\rm EO}=-6049$ (or $H/n=-1.7923$), a
significant improvement on the upper bound and already lower than
$\lim_{n\to\infty}H/n\approx1.786\ldots$ found in
Refs.~\cite{Pal,Hartmann,eo_prl}. Furthermore, we collected $10^5$
such states, which roughly segregate into 3 clusters with a mutual
Hamming distance of at least 100 distinct spins; at best a small
sample of the $\approx10^73$ ground states expected \cite{Hartmann_entro}!
For the smaller instance the bounds given are -922 and -912, resp.,
while EO finds -916 (or $H/n=-1.7891$) and was terminated after
finding $10^5$ such states. While this run (including sampling
degenerate states) took only a few minutes of CPU (at 800MHz), the
results for the larger instance required about 16 hours.

Finally, we note that we have considered the algorithm for making
Bethe lattices previously in Refs.~\cite{eo_jam,eo_perc}. In
Ref.~\cite{eo_perc} we have studied the graph bipartitioning problem
and found that the ground state energy was well above previous RS
calculations from Ref.~\cite{MP}, but only minutely below numerical
calculations obtained using simulated annealing~\cite{Banavar}. In
Ref.~\cite{eo_jam} we have considered some variations in the
generation of Bethe lattices and found that they effect the results
only in next-to-leading order.

\section{Numerical Results for Bethe Lattices}
\label{numerics}
We have simulated Bethe lattices with the algorithm described in
Sec.~\ref{EOalgo} for $k+1$ between 3 and 26, and graph sizes $n=2^l$
for $l=5,6,\ldots,10$ to obtain results for ground state energies, and
for $n\in[16\ldots256]$ to determine their entropy. In the following,
we present the results for ground-state energies and entropies from
those simulations. The results are discussed in detail in
Sec.~\ref{discussion}.

\subsection{Ground State Energies}
\label{GSE}
To reach relative statistical errors of our averages roughly uniform with
$n$ we generated initially a number of $10^5/\sqrt{n}$ instances for
each $n$ and $k+1$. Fortunately, deviations appear to narrow much
faster than $1/\sqrt{n}$, and thus we added more instances at smaller
$n$ with small extra computational cost to obtain narrow error bars
there as well. In Tab.~\ref{alldata} we list the values of average energies
according to Eq.~(\ref{eeq}), $\left<e_{k+1}(n)\right>$, for each
$k+1$ and $n$, the number of instances used, and the average number of
update steps required. Tab.~\ref{EOdata} lists a few properties of the 
computations. The results for the number of updates has
been also averaged over all connectivities $k+1$, although
lower-connected graphs require typically fewer updates. Note that this
is the {\it minimal} number of updates needed to obtain the listed
results, the actual number of updates taken up by each run of EO to
ensure convergence was at least twice of that but could be much
larger, according to the specification of the algorithm in
Sec.~\protect\ref{EOalgo}.

\begin{table*}
\caption{Data from the EO simulations for the average ground-state
energy per spin $e_{k+1}(n)$, plotted also in
Figs.~\protect\ref{3extraplot} and~\protect\ref{extrapolationplot}.}
\begin{tabular}{rlllllllllll}
\hline\hline
$n$ &  $-e_3(n)$ & $-e_4(n)$ & $-e_5(n)$ & $-e_6(n)$ & $-e_7(n)$ & $-e_8(n)$ & $-e_9(n)$ & $-e_{10}(n)$ & $-e_{15}(n)$ & $-e_{20}(n)$ & $-e_{25}(n)$  \\
\hline   
32 & 1.3506(6) &  1.5543(6) &  1.6734(8) &  1.8424(8) &  1.9425(9) & 2.0906(9) &  2.1730(33) &  2.7013(40) &  3.1056(48) &  3.4923(51)  \\
64 &1.2231(3) &  1.3964(4) &  1.5979(4) &  1.7294(5) &  1.8972(5) & 2.0083(6) &  2.1551(6) &  2.2557(23) &  2.7884(28) &  3.2090(33) & 3.6029(37) \\
128 &1.2426(5) &  1.4245(10) &  1.6269(10) &  1.7652(12) & 1.9335(13) & 2.0476(14) &  2.1945(15) &   2.3042(16) &   2.8443(20) &   3.2742(23) & 3.6731(25) \\
256 &1.2542(3) &  1.4417(6) &  1.6434(7) &  1.7885(8) &  1.9549(8) & 2.0782(10) &  2.2204(10) &   2.3324(11) &   2.8774(14) &   3.3186(16) & 3.7215(18) \\
512 &1.2608(2) &  1.4534(4) &  1.6548(5) &  1.8020(5) &  1.9685(6) & 2.0934(6) &  2.2379(7) &   2.3488(7) &   2.8993(10) &   3.3435(11) & 3.7505(12) \\
1024 &1.2644(1) &  1.4603(3) &  1.6612(3) &  1.8110(3) &  1.9762(5) & 2.1035(5) &  2.2470(5) &   2.3605(5) &   2.9092(7) &   3.3551(9) & 3.7612(11)
 \\
2048 &1.2673(1) &&&&&&&&&\\
\hline
$\infty$ &  1.2719(5) & 1.472(1)  & 1.673(1) & 1.826(1) & 1.990(3)  & 2.121(1)  &  2.2645(5)  & 2.378(3) &  2.935(1) &  3.389(1) & 3.806(4)\\ 
\hline\hline
\end{tabular}
\label{alldata}
\end{table*}

\begin{table}
\caption{Some properties of the numerical computations. Listed are for
each $n$ the number of instance used and the average number of updates
for each instance needed to obtain the results listed in
Tab.~\protect\ref{alldata}.}
\begin{tabular}{rcrcc}
\hline\hline
$n$ &~~~~~& Instances &~~~~~& $t$ \\
\hline   
32 && 19444 &&3.0\,$10^2$ \\
64 && 13750 &&1.5\,$10^3$\\
128 && 883 &&1.0\,$10^4$\\
256 && 625 &&1.6\,$10^5$\\
512 && 441 &&3.1\,$10^6$\\
1024 && 312 &&7.6\,$10^7$\\
2048 && 220 && 1.5\,$10^8$\\
\hline\hline
\end{tabular}
\label{EOdata}
\end{table}

\begin{figure}
\vskip 2.6in  
\includegraphics{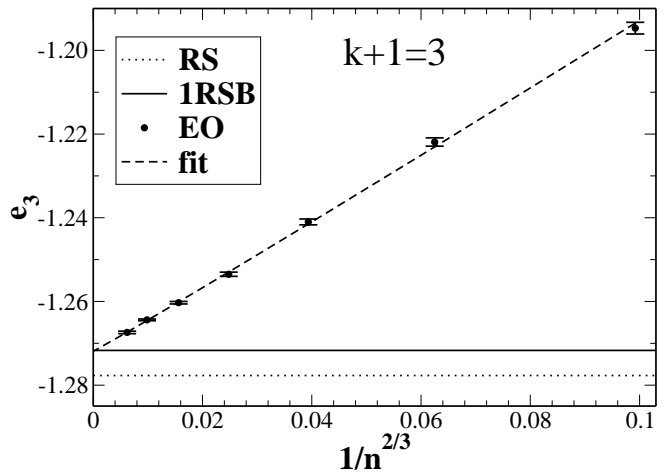}
\caption{Extrapolation plot for the EO data for $k+1=3$ given in
Tab.~\protect\ref{alldata} and the fitted curve according to
Eq.~(\protect\ref{extrapolationeq}).  (Data points were weighted with
respect to $n$ and the inverse of the error.)  For $n\to\infty$ the
extrapolation gives $e_3=-1.2719(5)$, way above the RS result but
consistent with the 1RSB result from Ref.~\protect\cite{MP2}, both
indicated by horizontal lines.  }
\label{3extraplot}
\end{figure}

\begin{figure*}
\vskip 8.5in  
\includegraphics{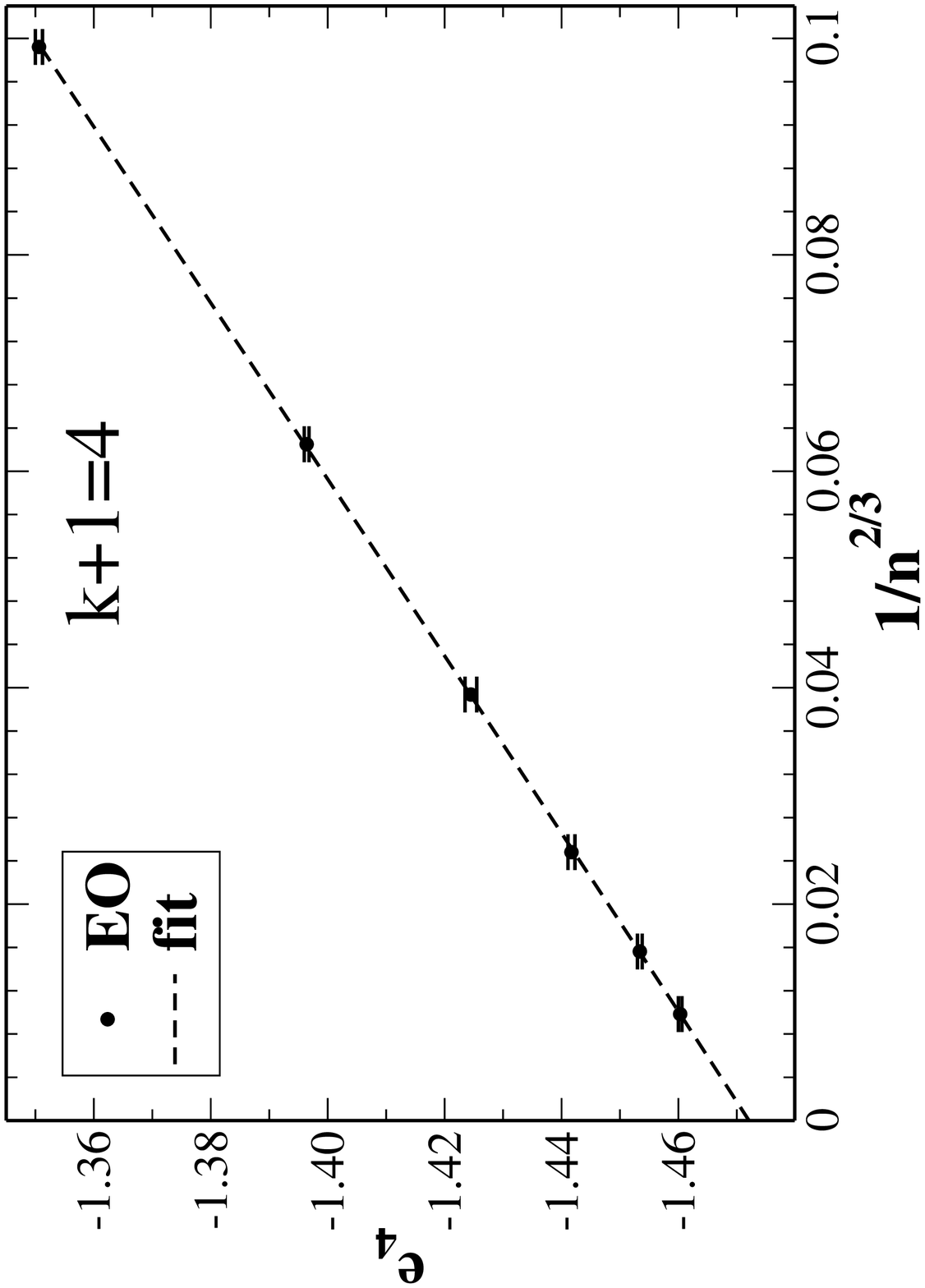} 
\includegraphics{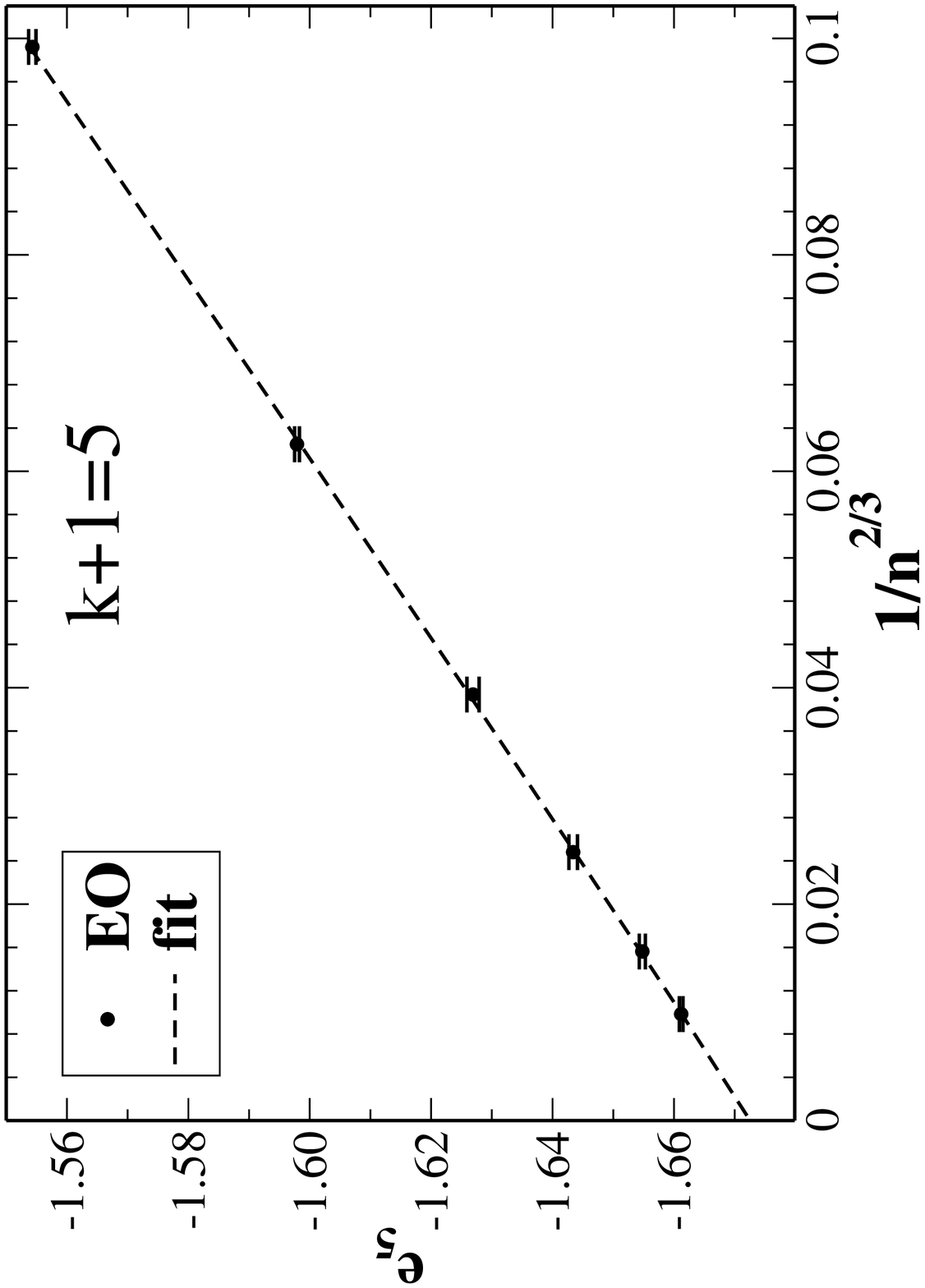}
\includegraphics{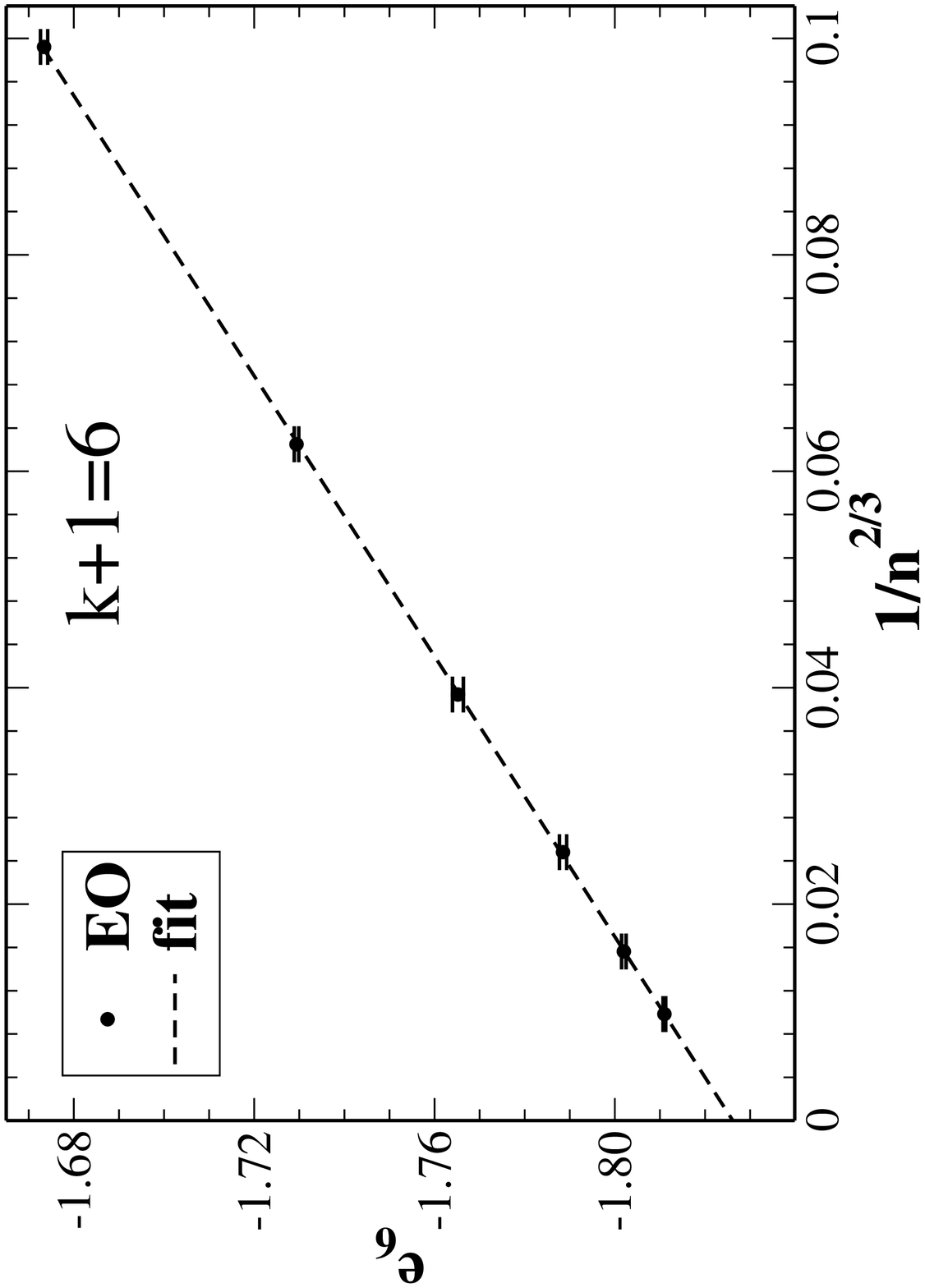} 
\includegraphics{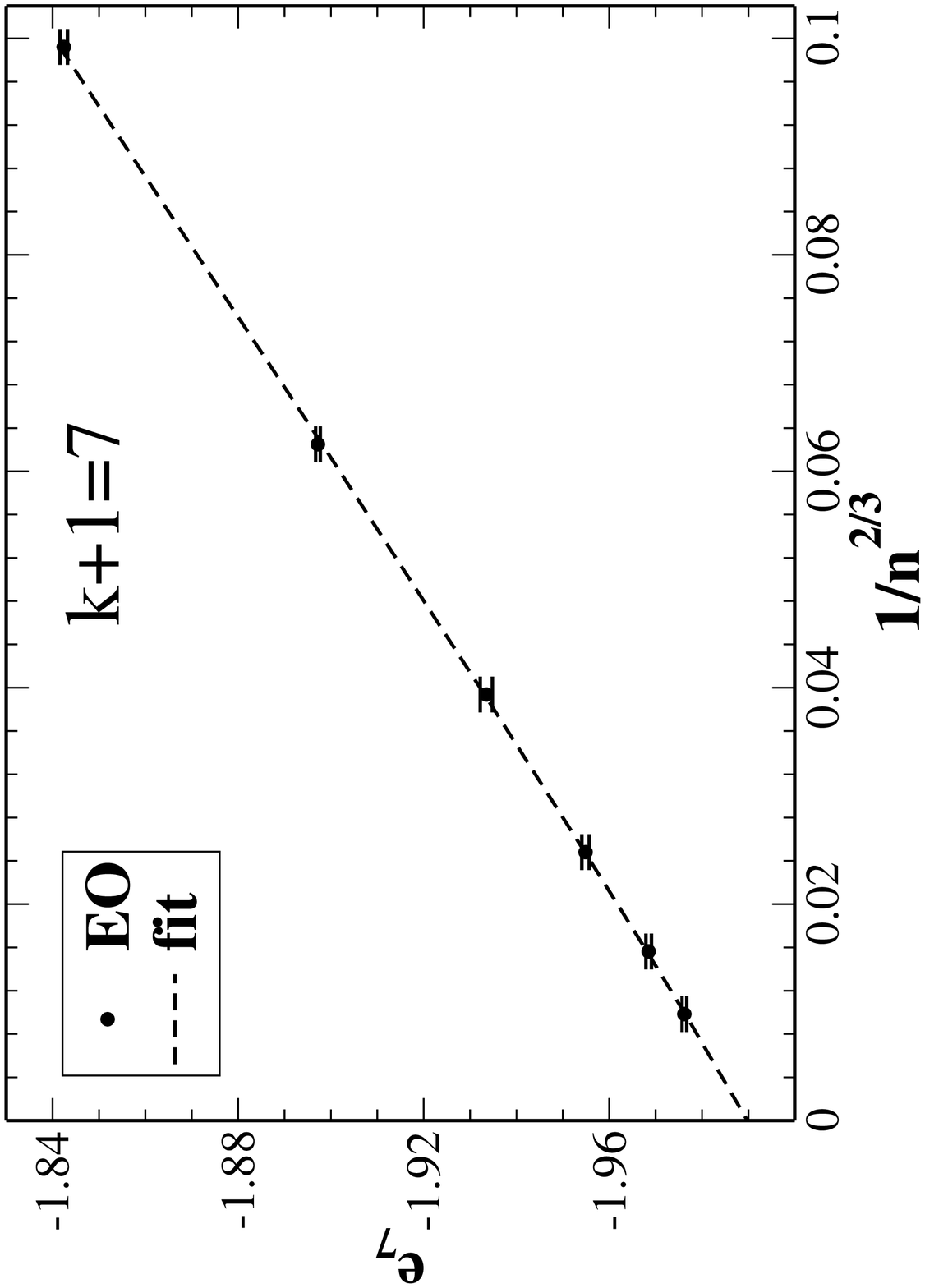}
\includegraphics{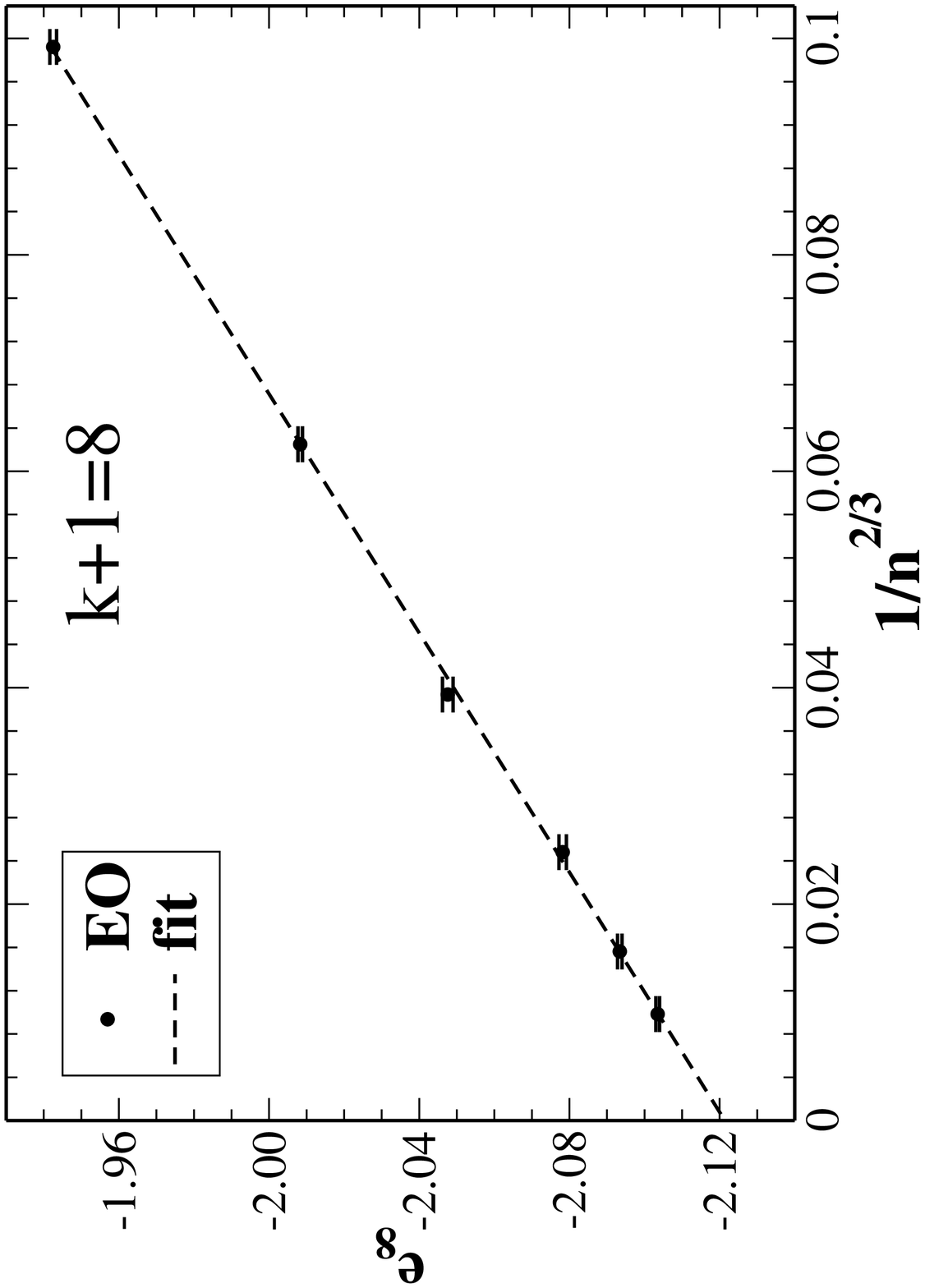} 
\includegraphics{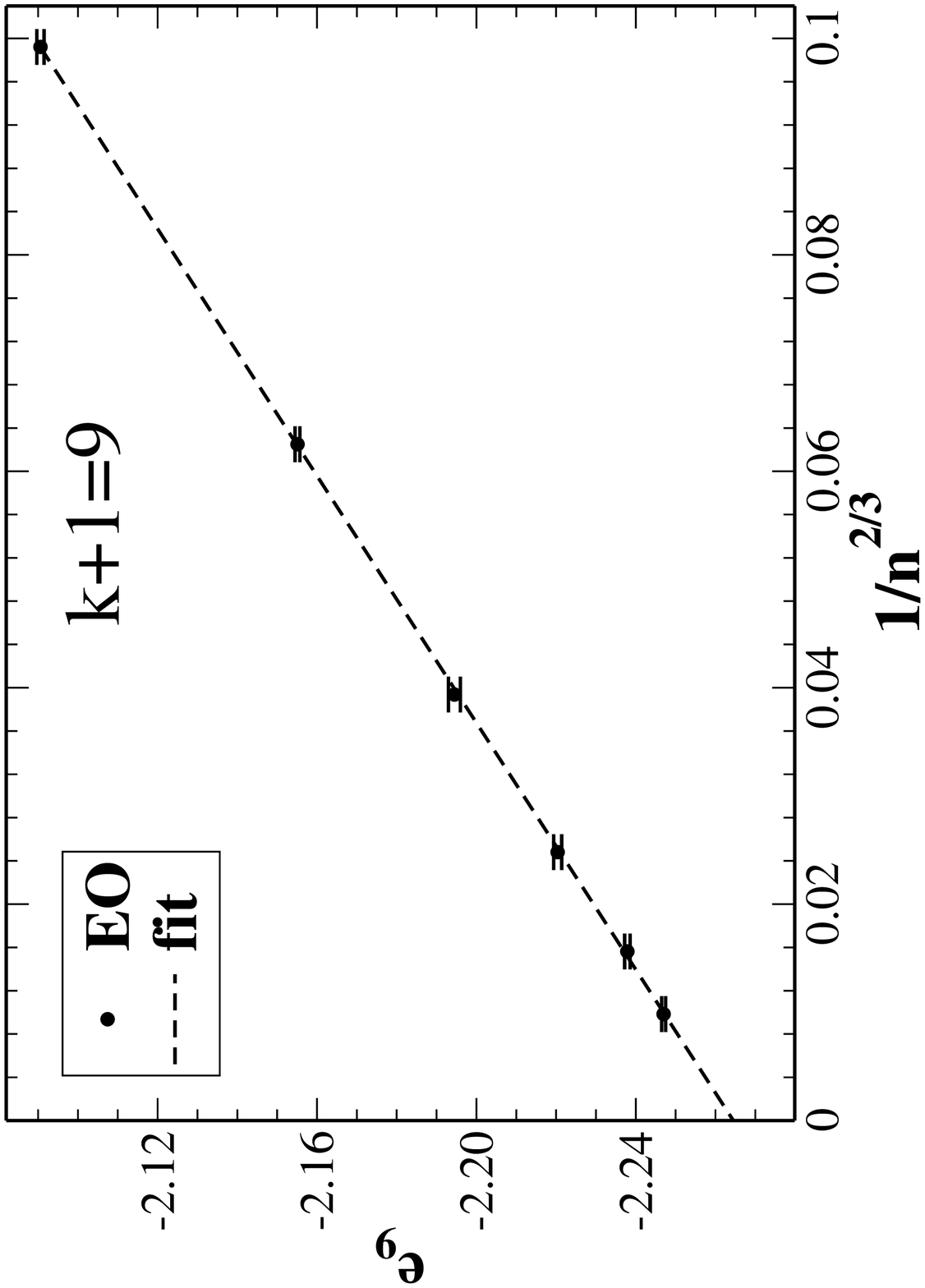}
\includegraphics{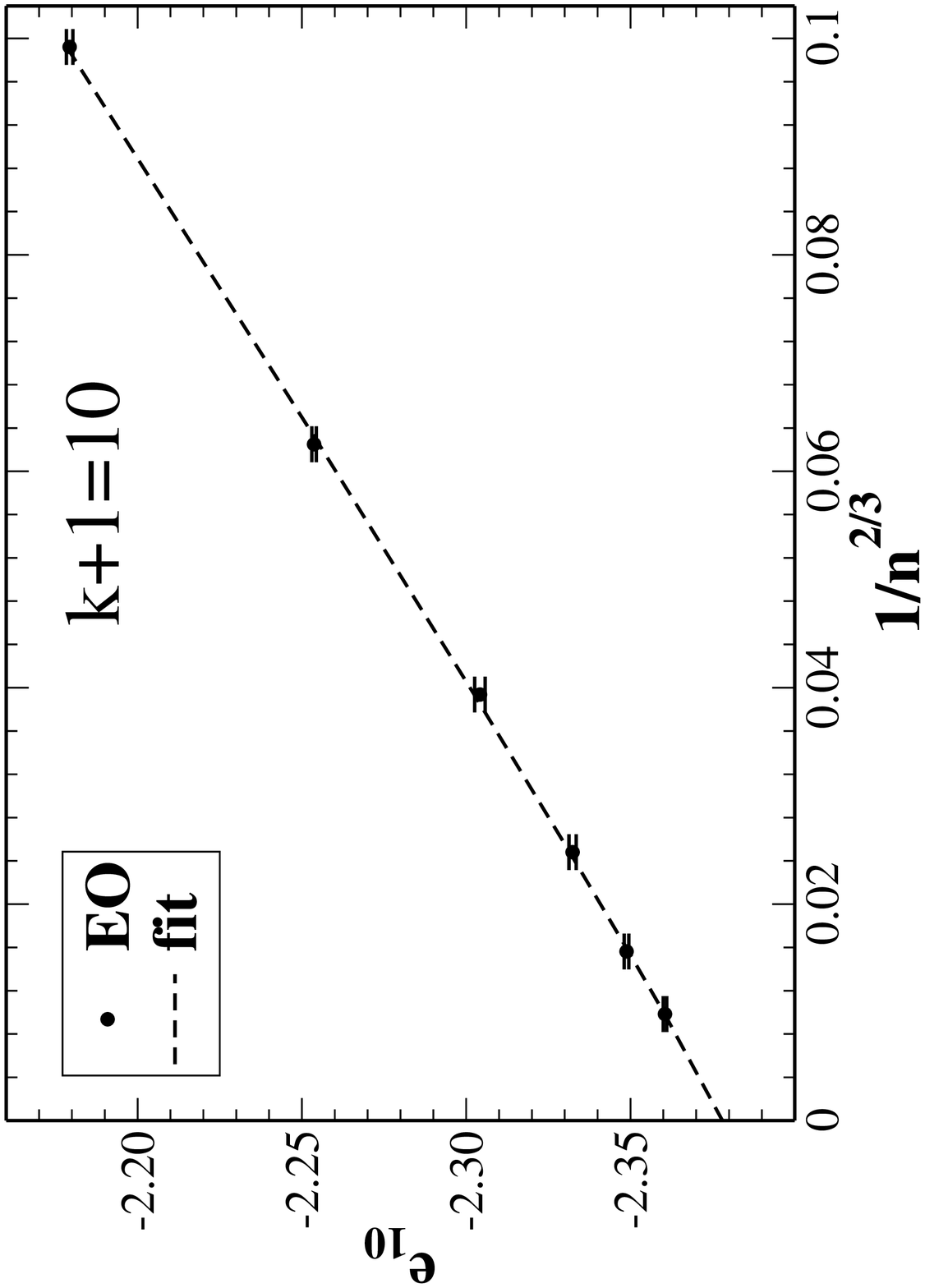} 
\includegraphics{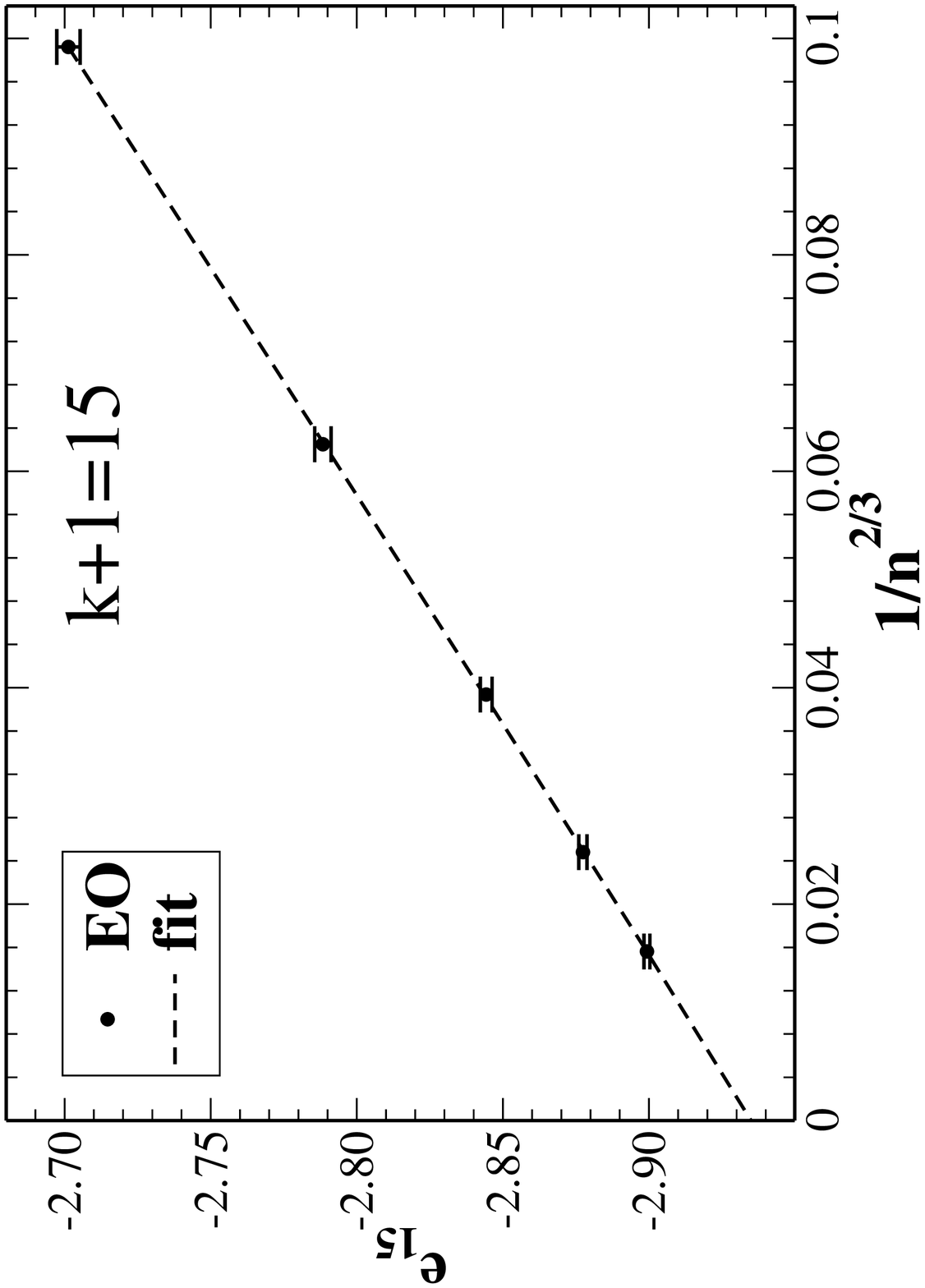}
\includegraphics{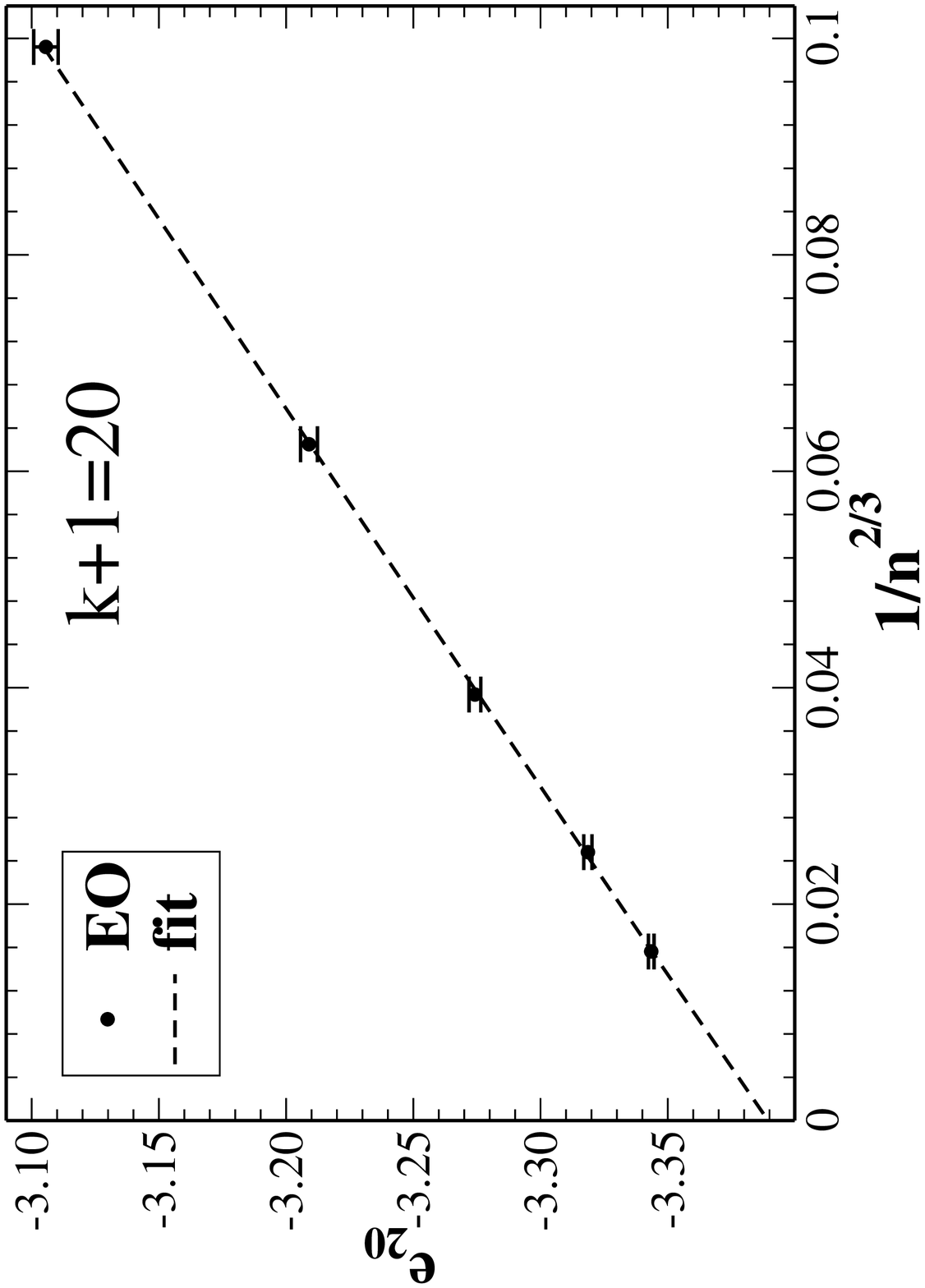} 
\includegraphics{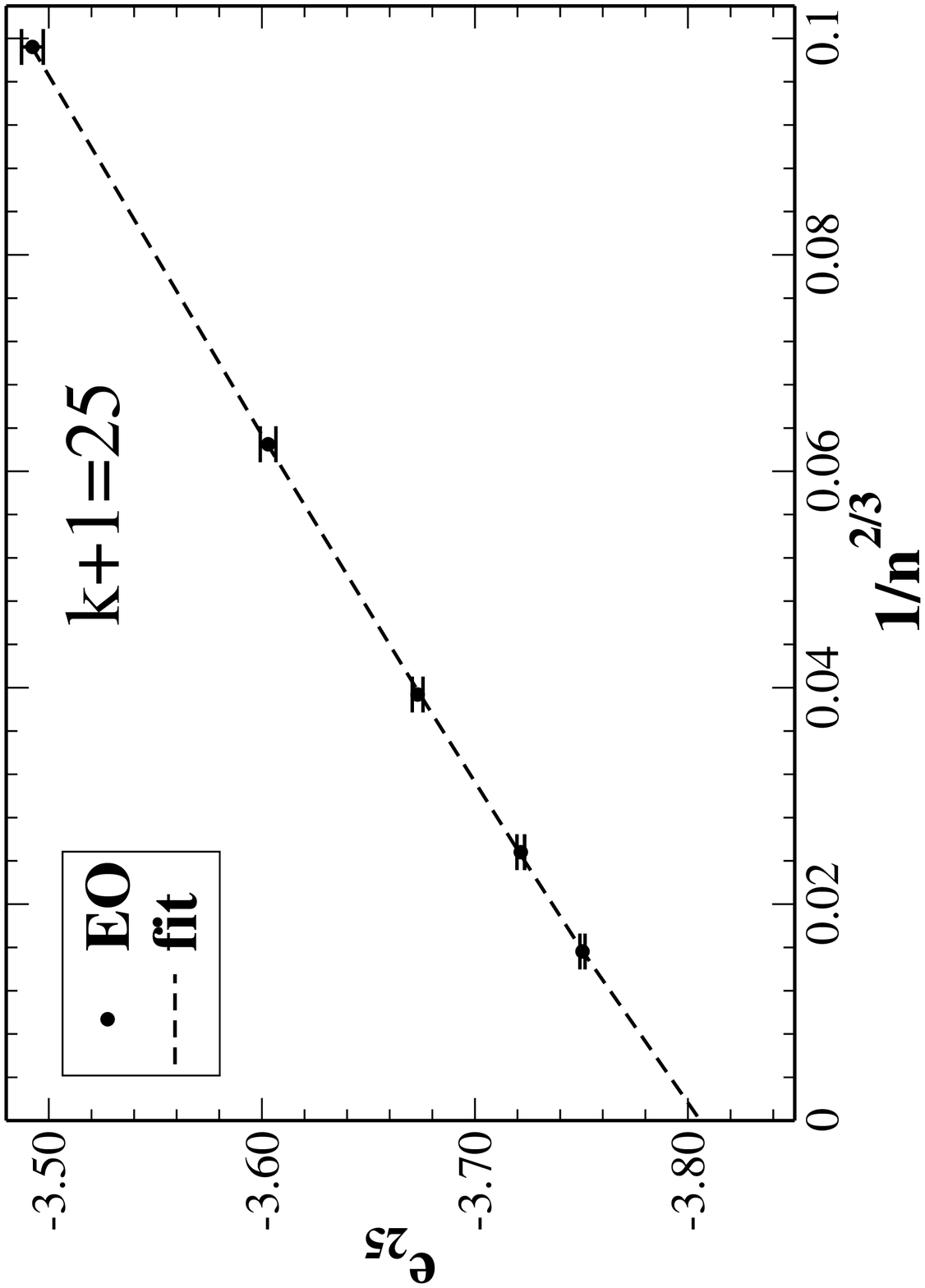}
\caption{Extrapolation plot for the EO data in
Tab.~\protect\ref{alldata} for $k+1=4$ to 25, as in
Fig.~\protect\ref{3extraplot}. All data seems to extrapolate well
linearly in $1/n^{2/3}$. The extrapolated values of $e_{k+1}$ for
$n\to\infty$ are also listed in Tab.~\protect\ref{alldata}.  }
\label{extrapolationplot}
\end{figure*}

Unfortunately, when plotted as a function of $1/n$, the average
energies for each given $k+1$ clearly do not extrapolate linearly (as,
for example, seems to be the case for spin glasses on a hypercubic
lattice \cite{Pal,Hartmann,eo_prl}). Instead, using an extrapolation
according to \footnote{In a previous simulation \cite{eo_jam}, we have
attempted to fit those data (at $k=1=3,4$) with a $\ln(n)/n$
correction with a result that was somewhat above the current
extrapolation value.}
\begin{eqnarray}
e_{k+1}(n)\sim e_{k+1}+\frac{A}{n^\nu}\quad(n\to\infty).
\label{extrapolationeq}
\end{eqnarray}
We find that for the whole range of connectivities $k+1$ studied here,
the scaling corrections appeared to be consistent with $\nu=2/3$
within a few percent, except for two outliers at $k+1=10$ and
25. Thus, we have plotted for each $k+1$ the values of $e_{k+1}(n)$ as
a function of $1/n^{2/3}$ in Figs.~\ref{3extraplot}
and~\ref{extrapolationplot}. Although the extrapolation appears to be
linear on that scale for each $k+1$, we have fitted the data with the
more general form of Eq.~(\ref{extrapolationeq}).
(Fits were weighted according to $n$ and to the inverse of
the standard deviation for each point.) These fits are also shown as
dashed lines in each of the Figs.~\ref{3extraplot}
and~\ref{extrapolationplot}.

The extrapolation results for the ground state energies appear to be
quite stable under variation of the scaling form, for instance, when
fitting with fixed $\nu=2/3$ instead of
Eq.~(\ref{extrapolationeq}). We estimate that each has a relative
error of about <0.3\%. Exceptions to this estimate we have to grant
for the cases of $k+1=10$ and 25, in which case we also observe
significant differences to the $\nu=2/3$ corrections to scaling.

We can compare our results with existing theoretical predictions at
the RS and the 1RSB level at least for the case of $k+1=3$. For this
case, a recently published calculation \cite{MP2} yielded
$e_3=-1.2777$ at the RS, and $e_3=-1.2717$ at the 1RSB level (further
replica corrections are expected to be small). These values are also
indicated in Fig.~\ref{3extraplot}. Clearly, our result for
$k+1$ is consistent with the 1RSB results, but certainly inconsistent
with the RS result. Further 1RSB results for other values $k+1$ are
currently being calculated \cite{Mezard_pc}. We will discuss a more
detailed analysis of the extrapolated values of $e_{k+1}$ at
$n\to\infty$ in Sec.~\ref{discussion}.

\subsection{Ground State Entropy}
\label{GSS}
We have also used EO to sample the degeneracy $\Omega$ of the
lowest-energy states found. Due to the discrete nature of the energy
of the system, ground-states can be highly degenerate, and the
ground-state entropy per spin defined in Eq.~(\ref{seq}) may well be
non-vanishing for $n\to\infty$. While the search for a ground state of
an instance is certain to provide a rigorous upper bound to the actual
ground state energy, the search for the complete set of ground states
for an instance entails the risk of a two competing systematic
errors. (1) If EO misses to find the exact ground state, one is likely
to vastly over-count the degeneracy, since $\Omega$ is expected to rise exponentially with the energy above the ground state~\cite{Sibani}. (2) Even if EO
finds ground states, it may simply undercount $\Omega$, since such
states could be too far separated in configuration space. Therefore, we
have implemented EO with the settings described in Sec.~\ref{EOalgo},
which emphasize the desire for accuracy over computational
efficiency. Accordingly, we were bound to conduct a separate set of
simulations from those that determined the energies only. In these
simulations we focused on smaller system sizes of $n\leq256$ for
$k+1=3,\ldots,9$ and $10,14,\ldots,26$ only. The limit on $n$ for the
smaller $k+1$ is mostly dictated by avoiding system sizes at which
$\Omega$ typically exceeds $10^6$.

 As a test for the accuracy of our implementation, we have run the
simulation for $k+1=3$ twice on the exactly identical instances, using
different initial conditions and $n/5$ more updates in the second run:
The results, both for the energies and $\Omega$, were identical {\it
for each instance,} producing the same set of configurations
independent of the starting point of the search. We therefore assume
that systematic errors in our data are small and can be neglected.

\begin{figure*}
\vskip 8.0in  
\includegraphics{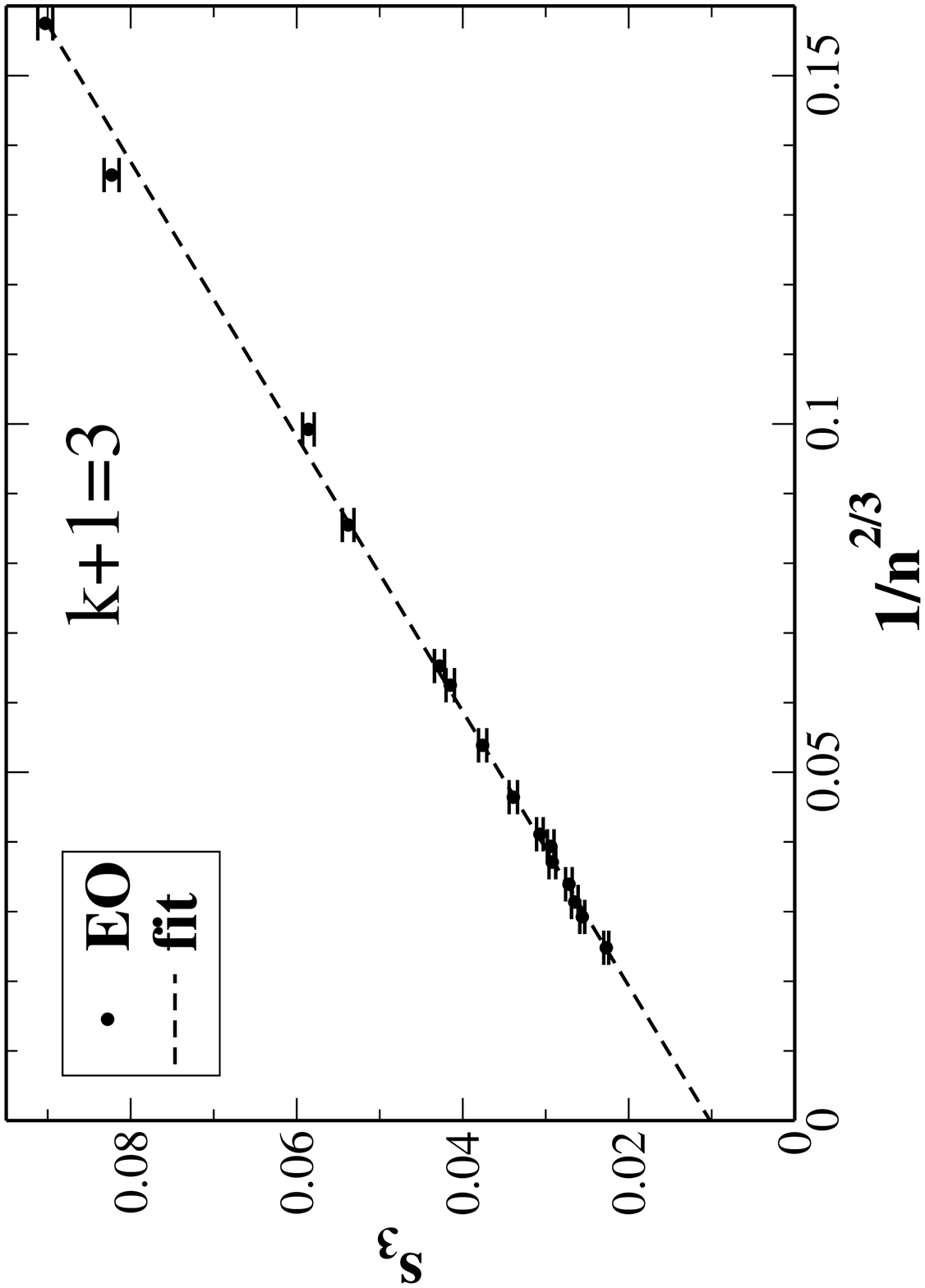} 
\includegraphics{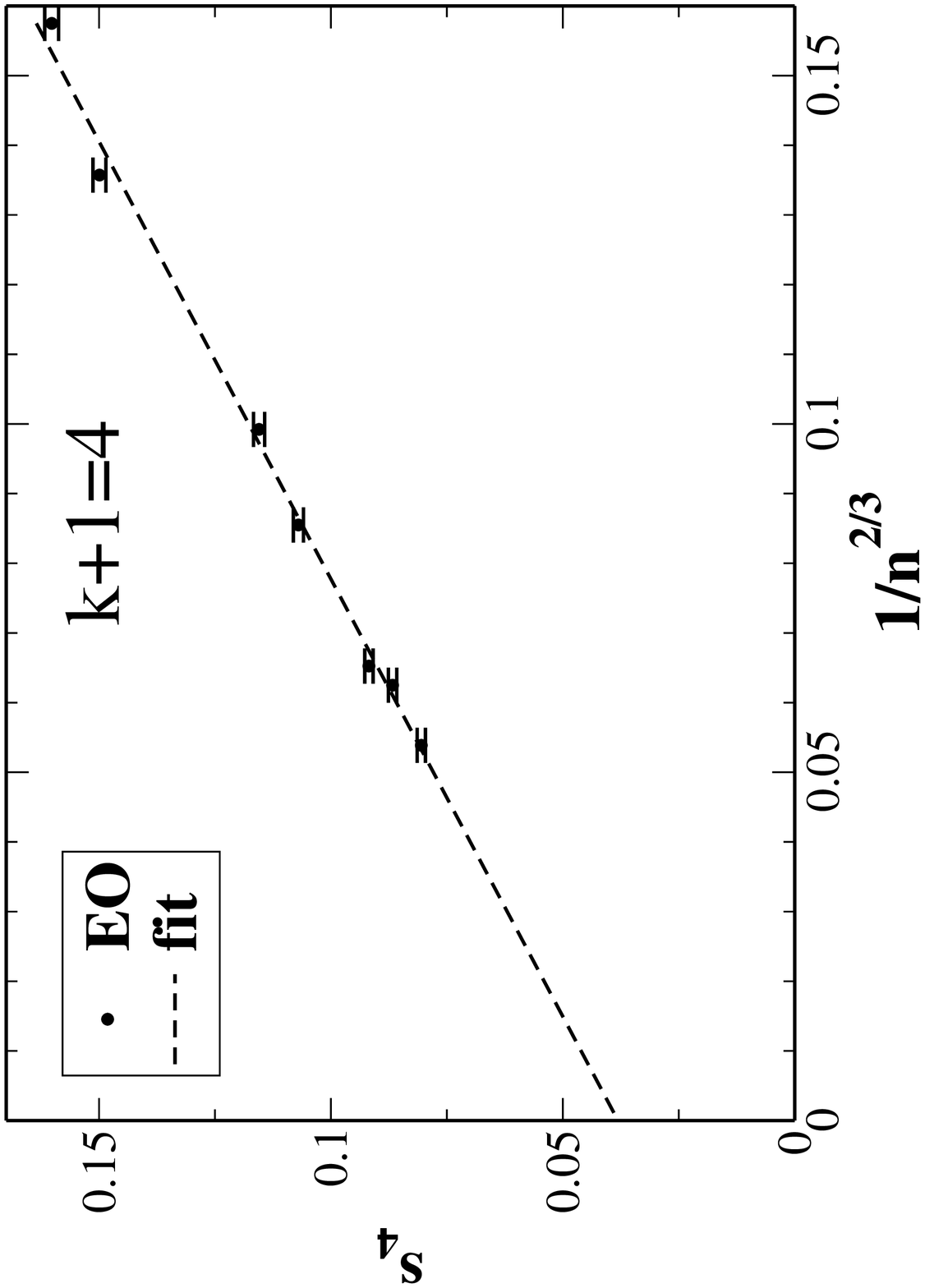}
\includegraphics{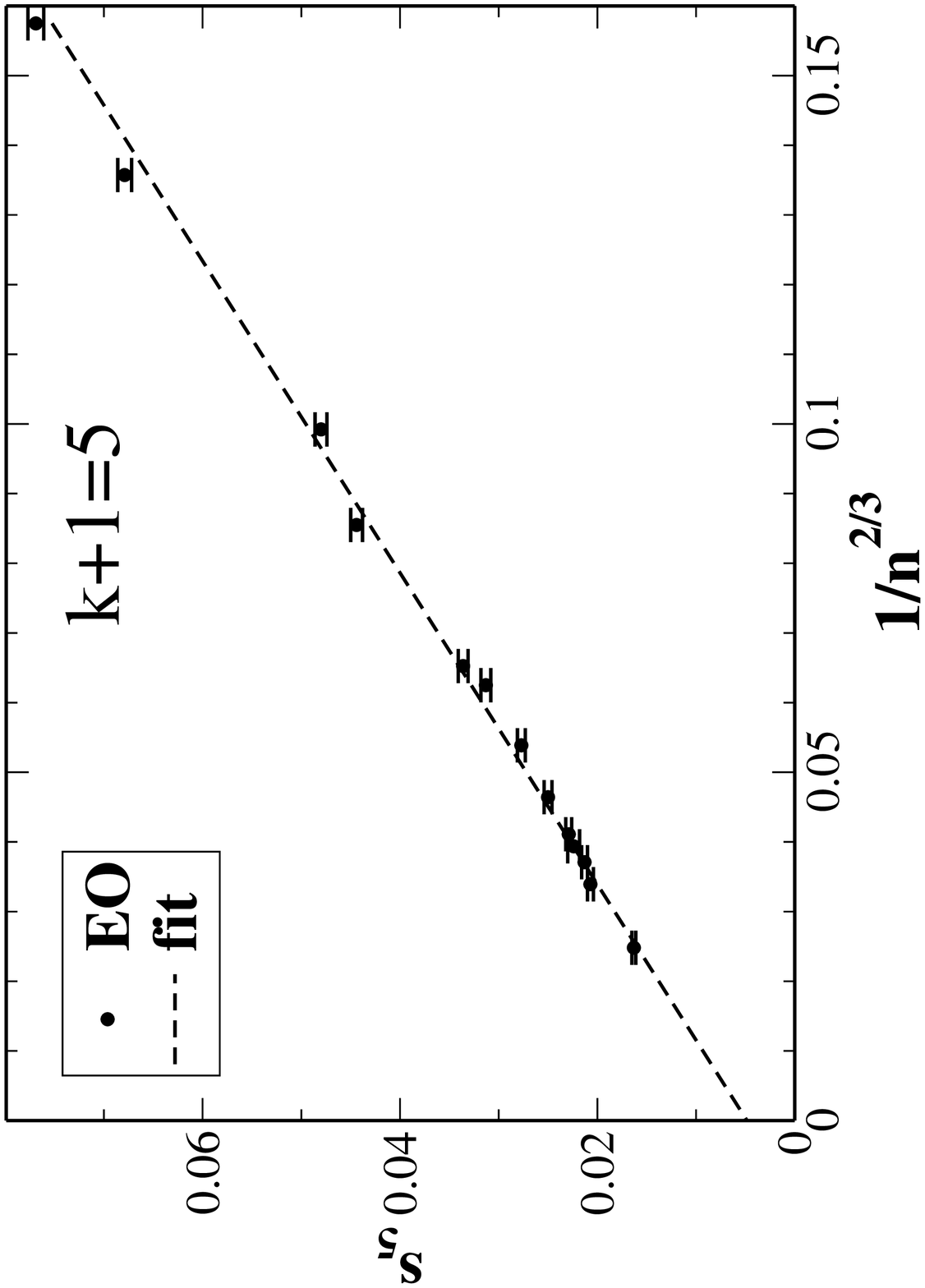} 
\includegraphics{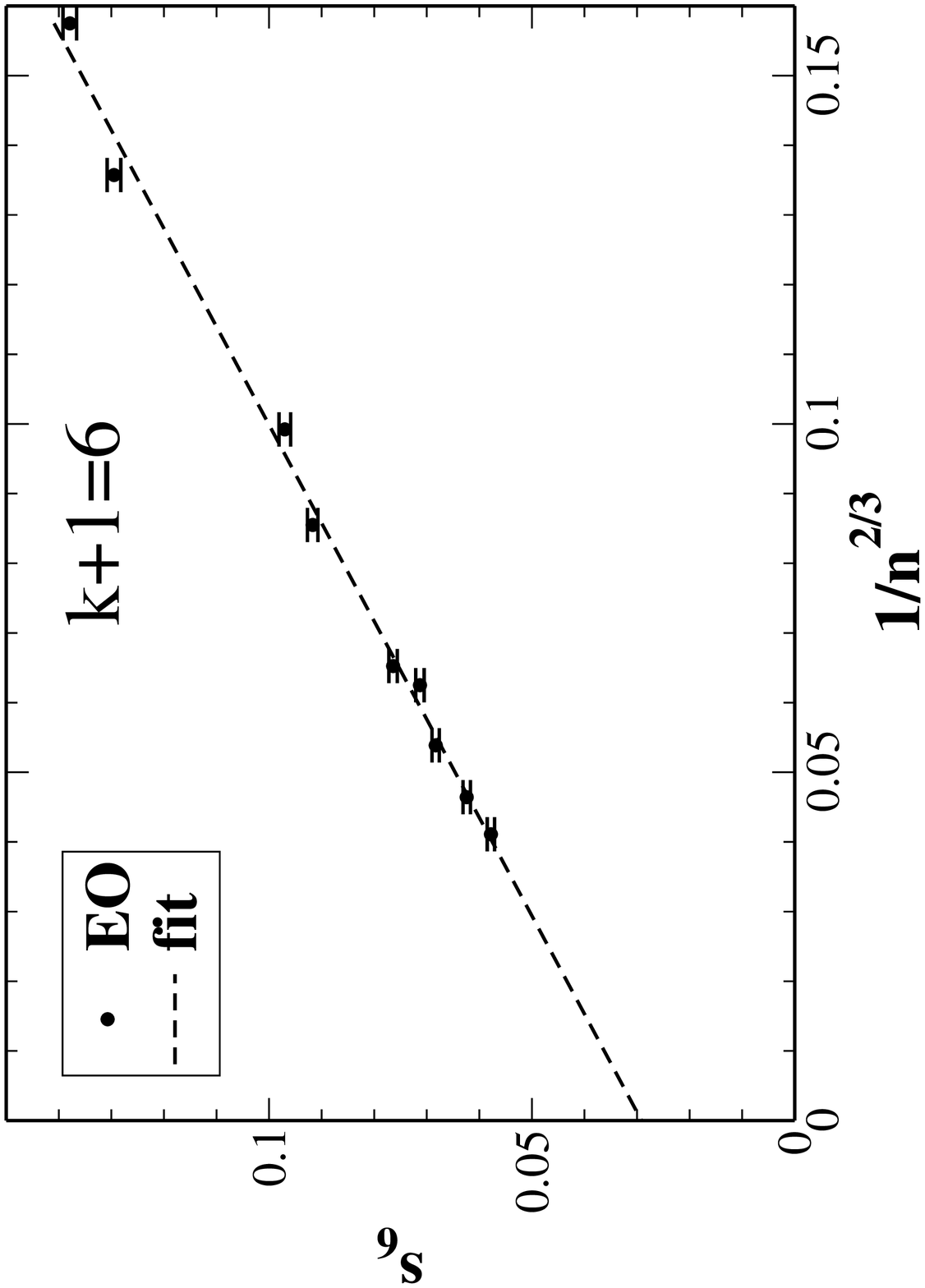}
\includegraphics{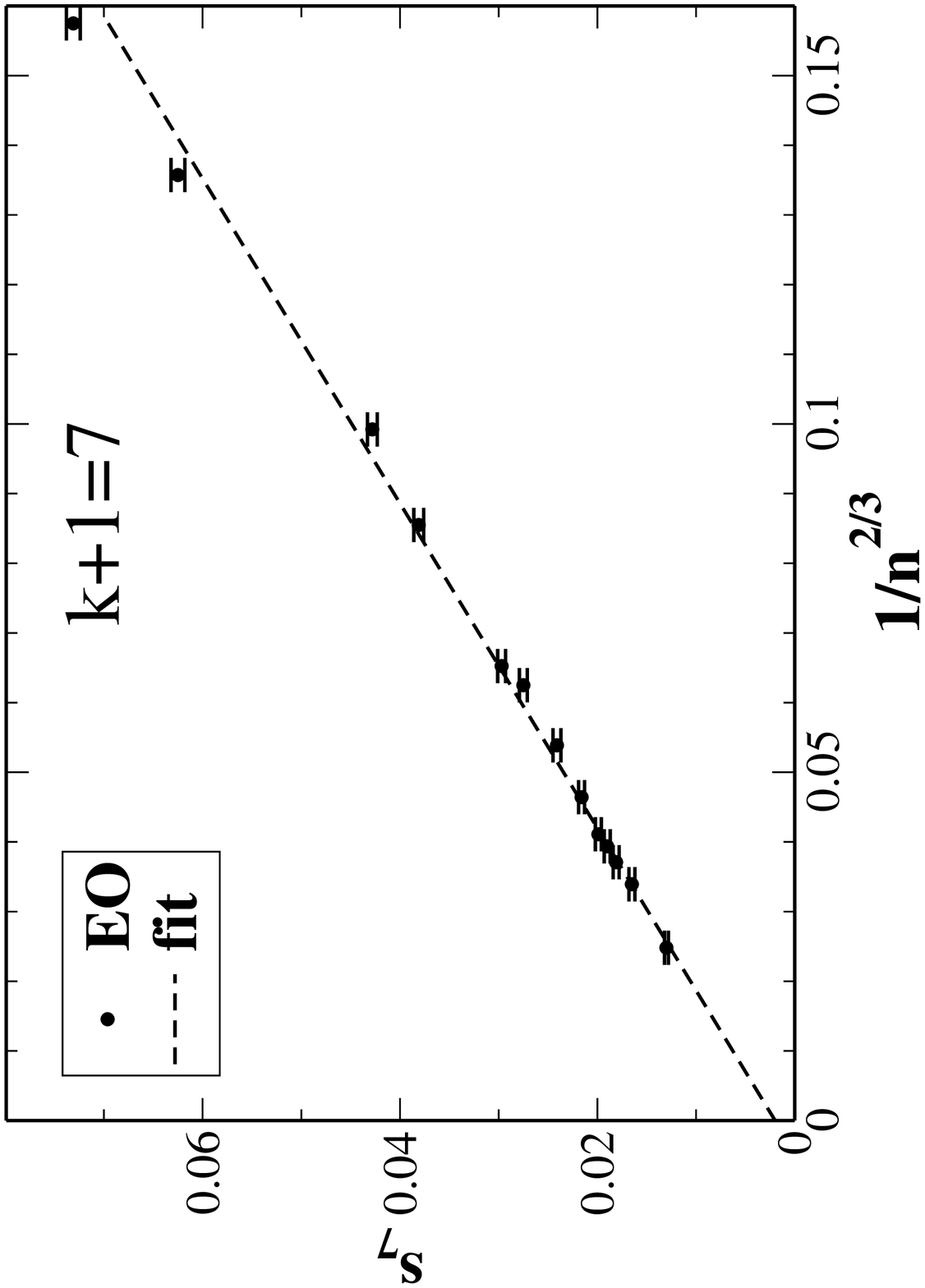} 
\includegraphics{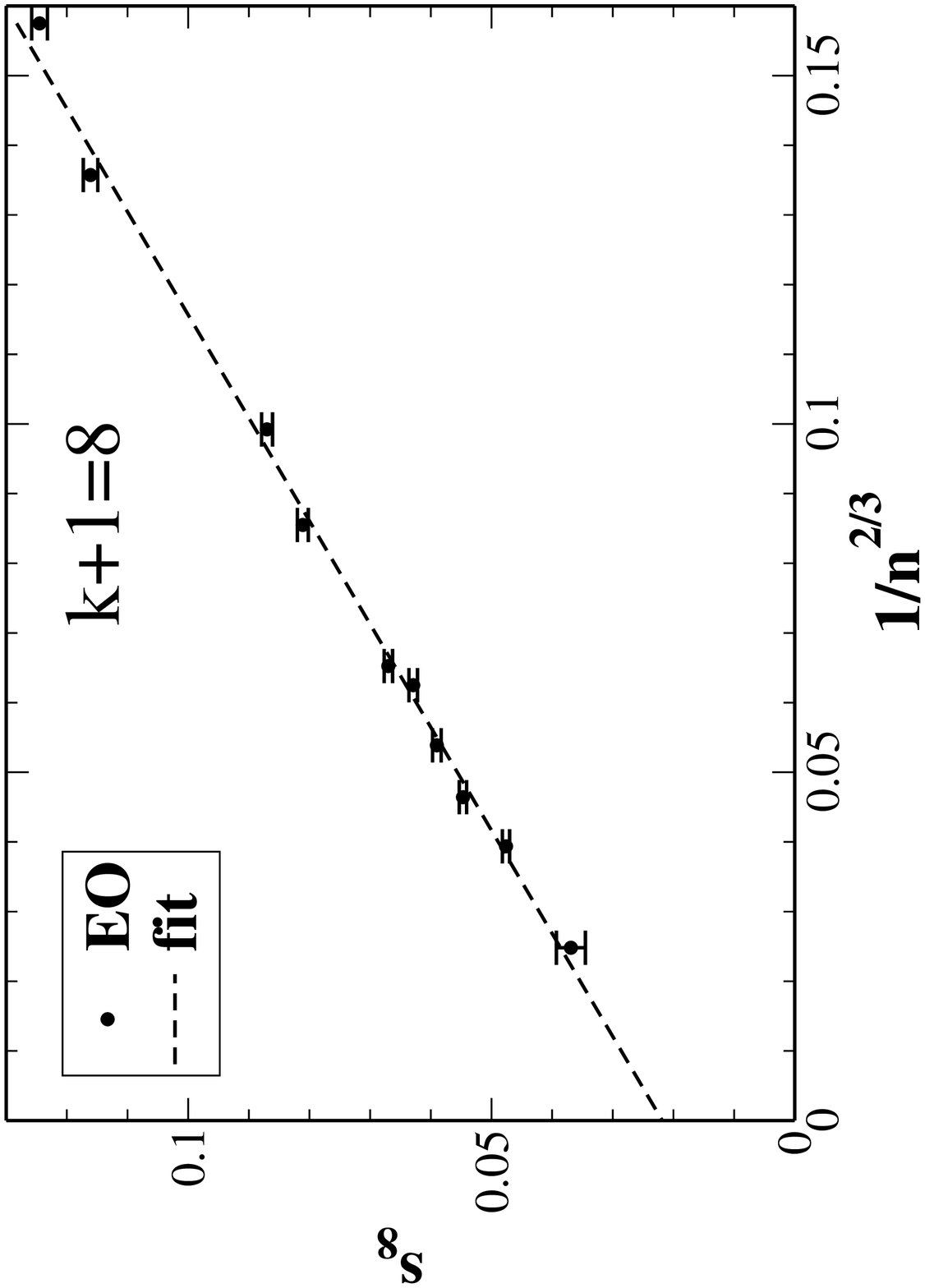}
\includegraphics{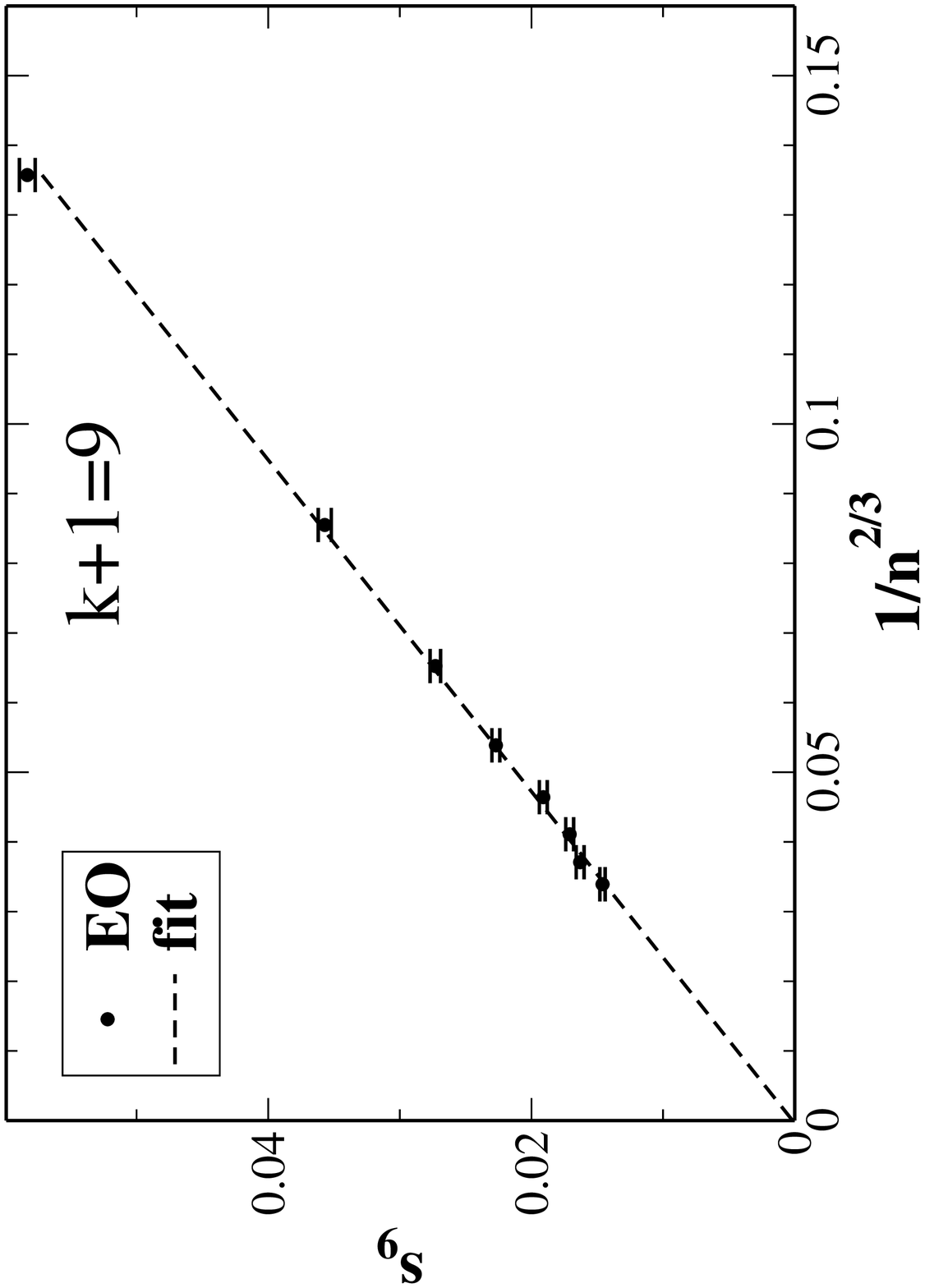} 
\includegraphics{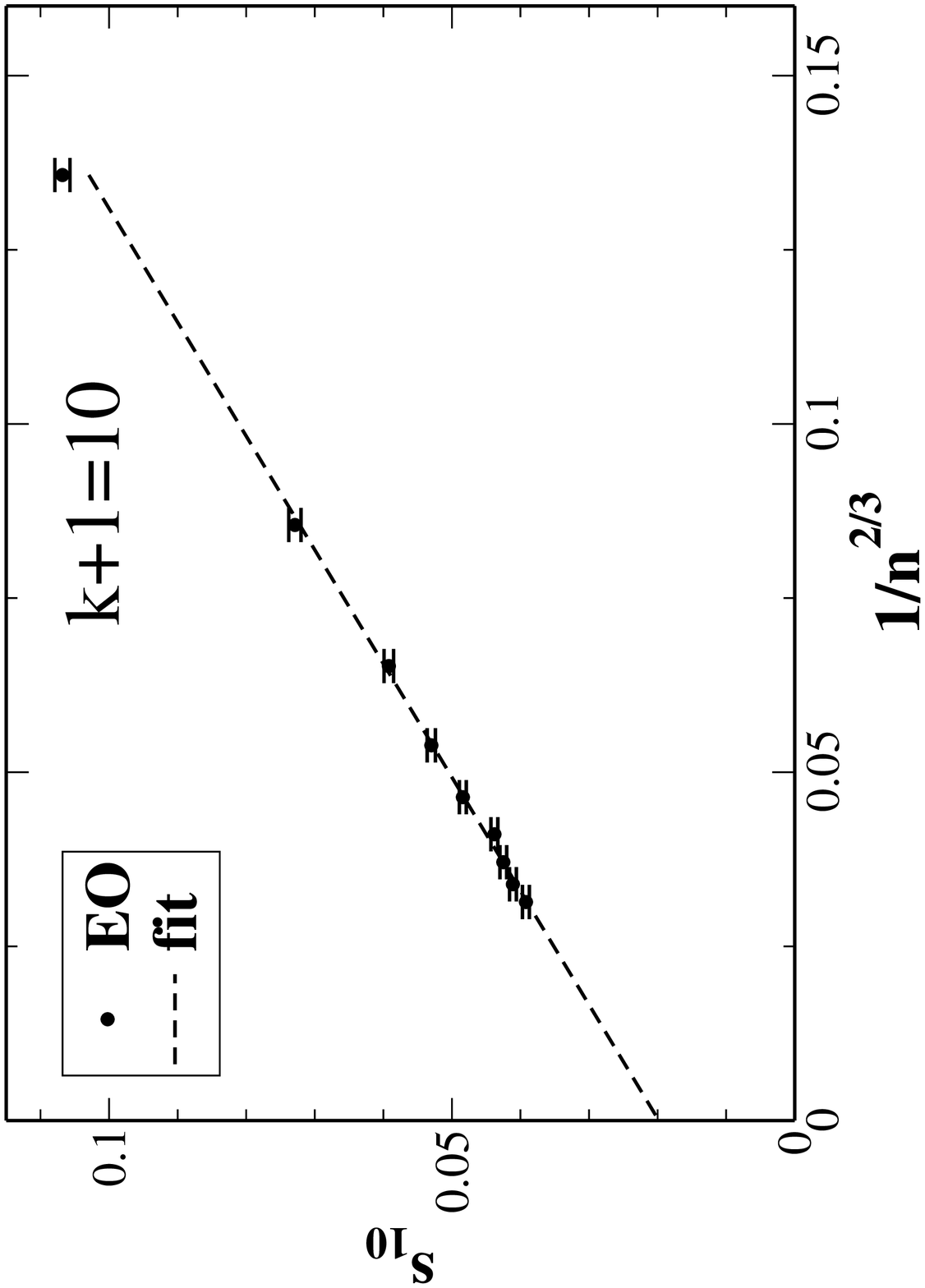}
\includegraphics{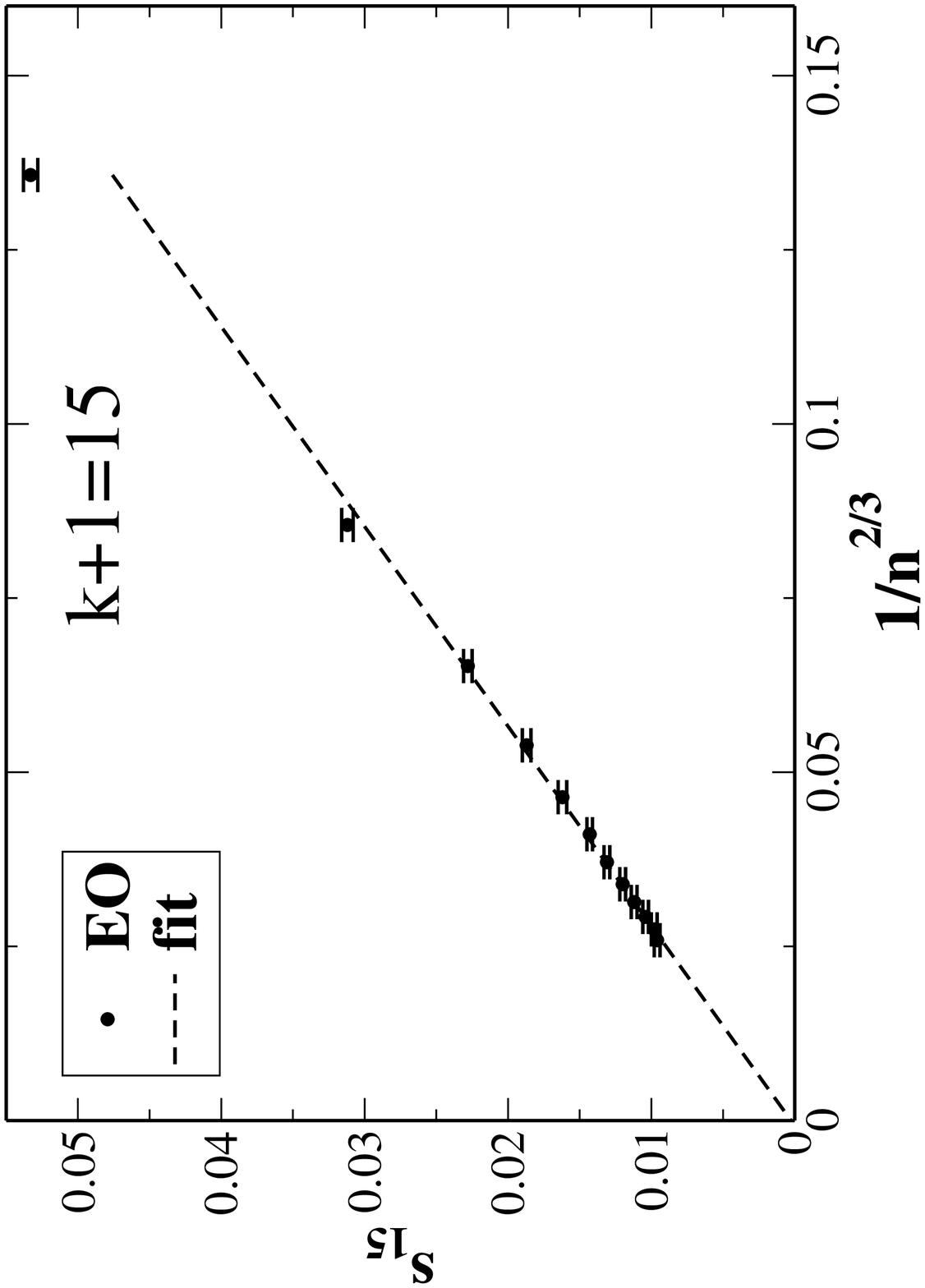} 
\includegraphics{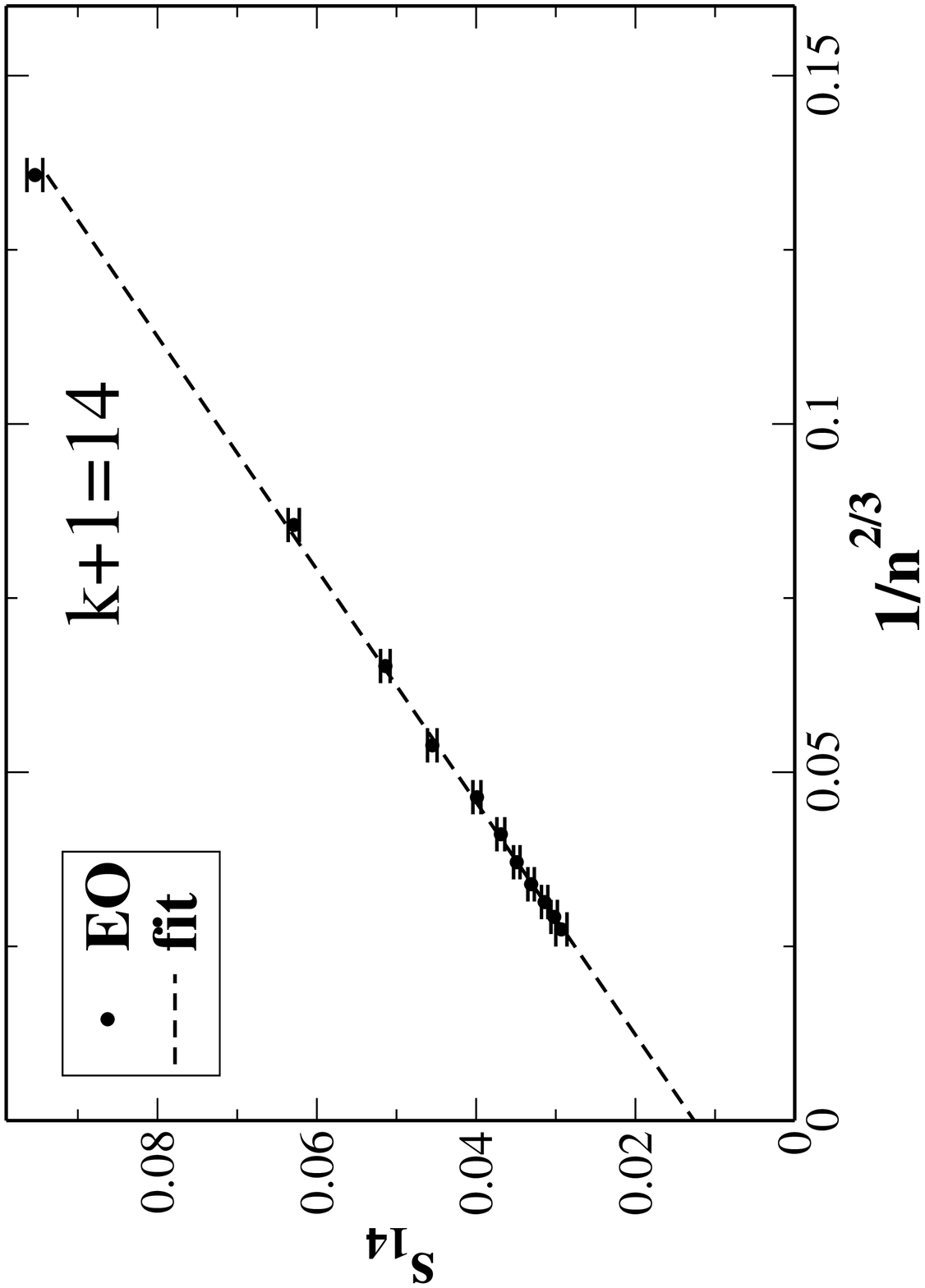}
\caption{Extrapolation plot for the EO data for the entropy $s_{k+1}(n)$ for $k+1=3$ to 15. All data seems to extrapolate well
linearly in $1/n^{2/3}$. Note the difference in the results between odd (left) and even (right) $k+1$. The extrapolated values of $s_{k+1}$ for
$n\to\infty$ are listed in Tab.~\protect\ref{entropy}.  }
\label{entroextraplot}
\end{figure*}

\begin{figure}
\vskip 5.3in
\includegraphics{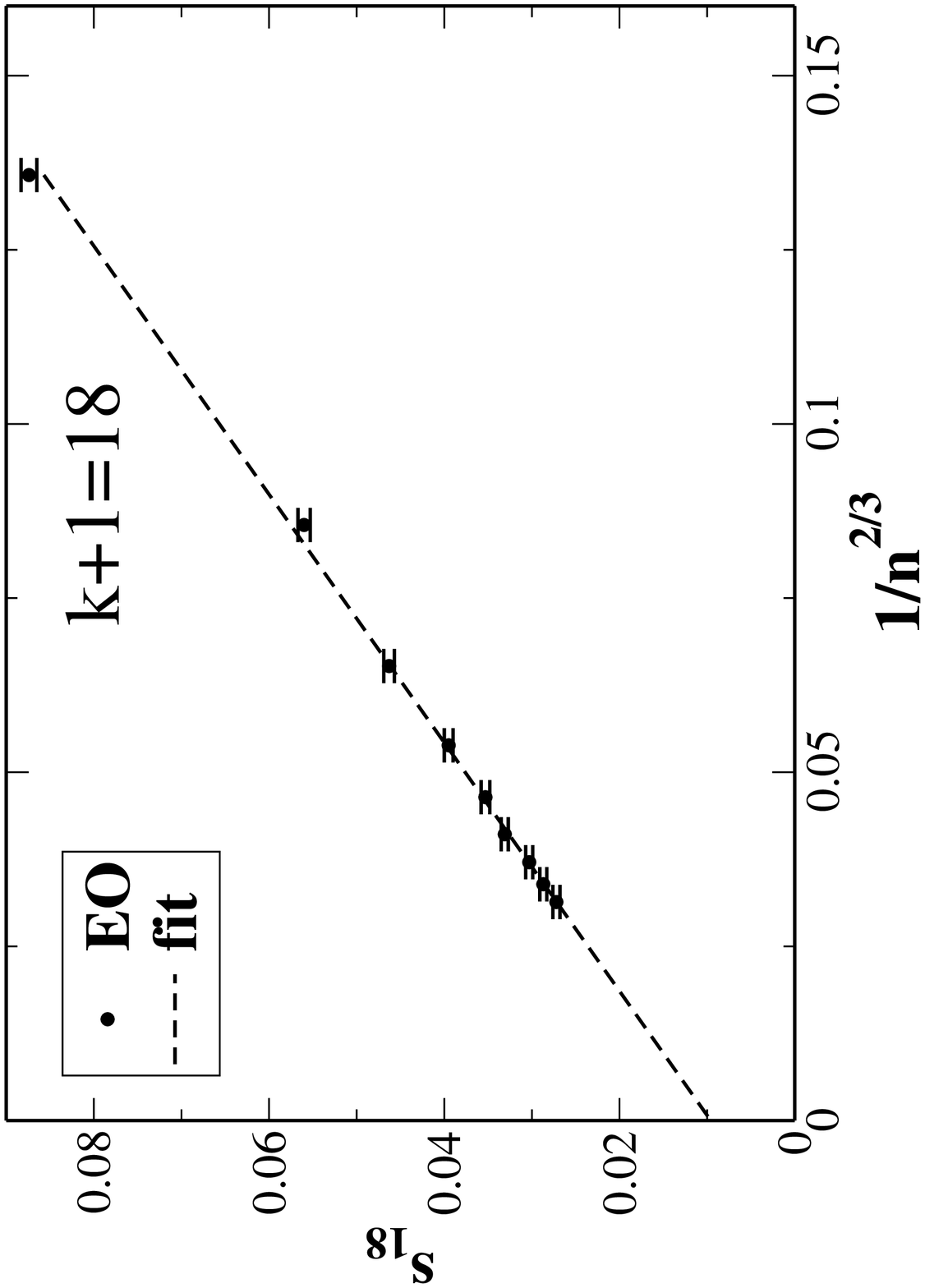} 
\includegraphics{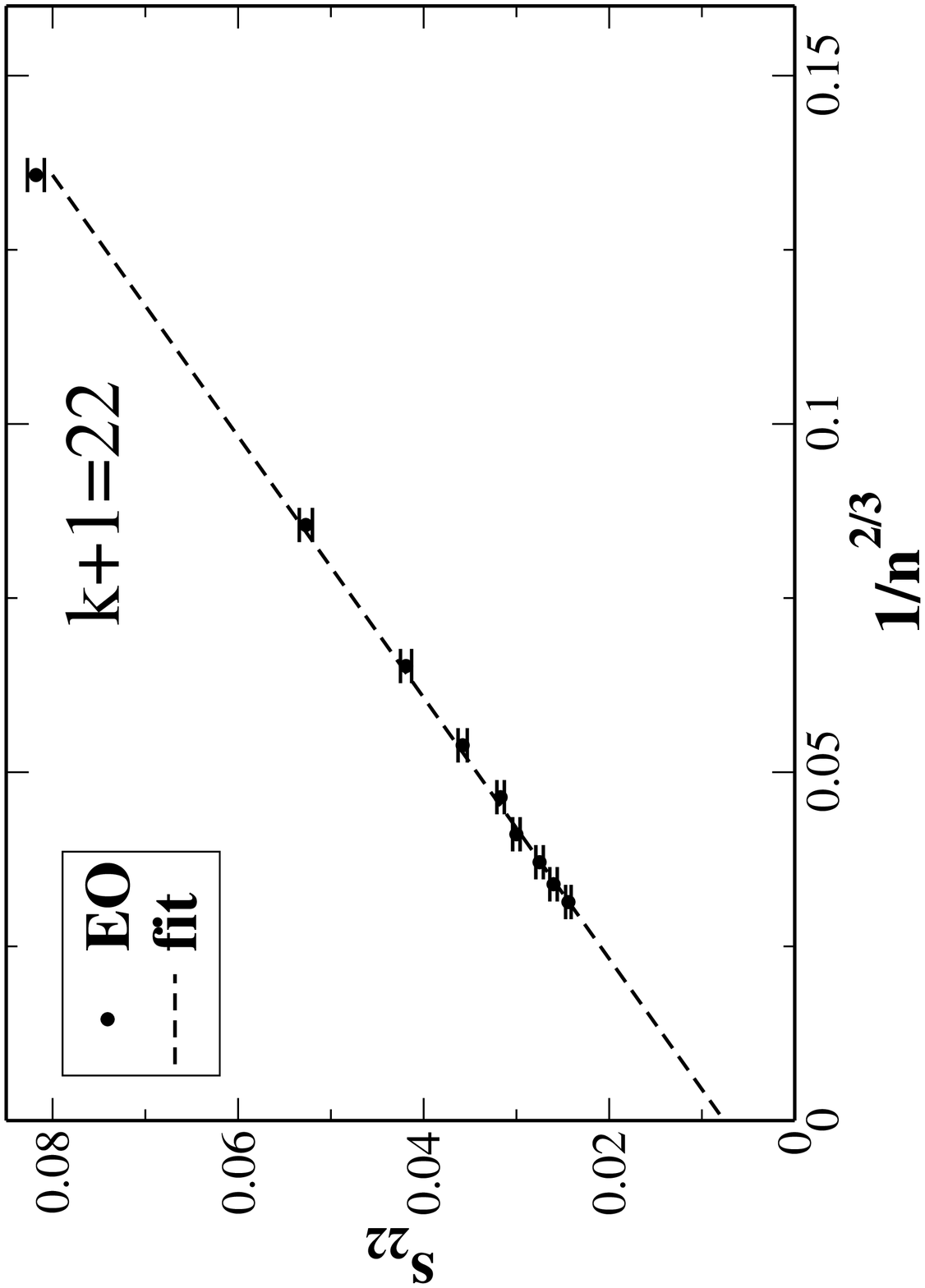} 
\includegraphics{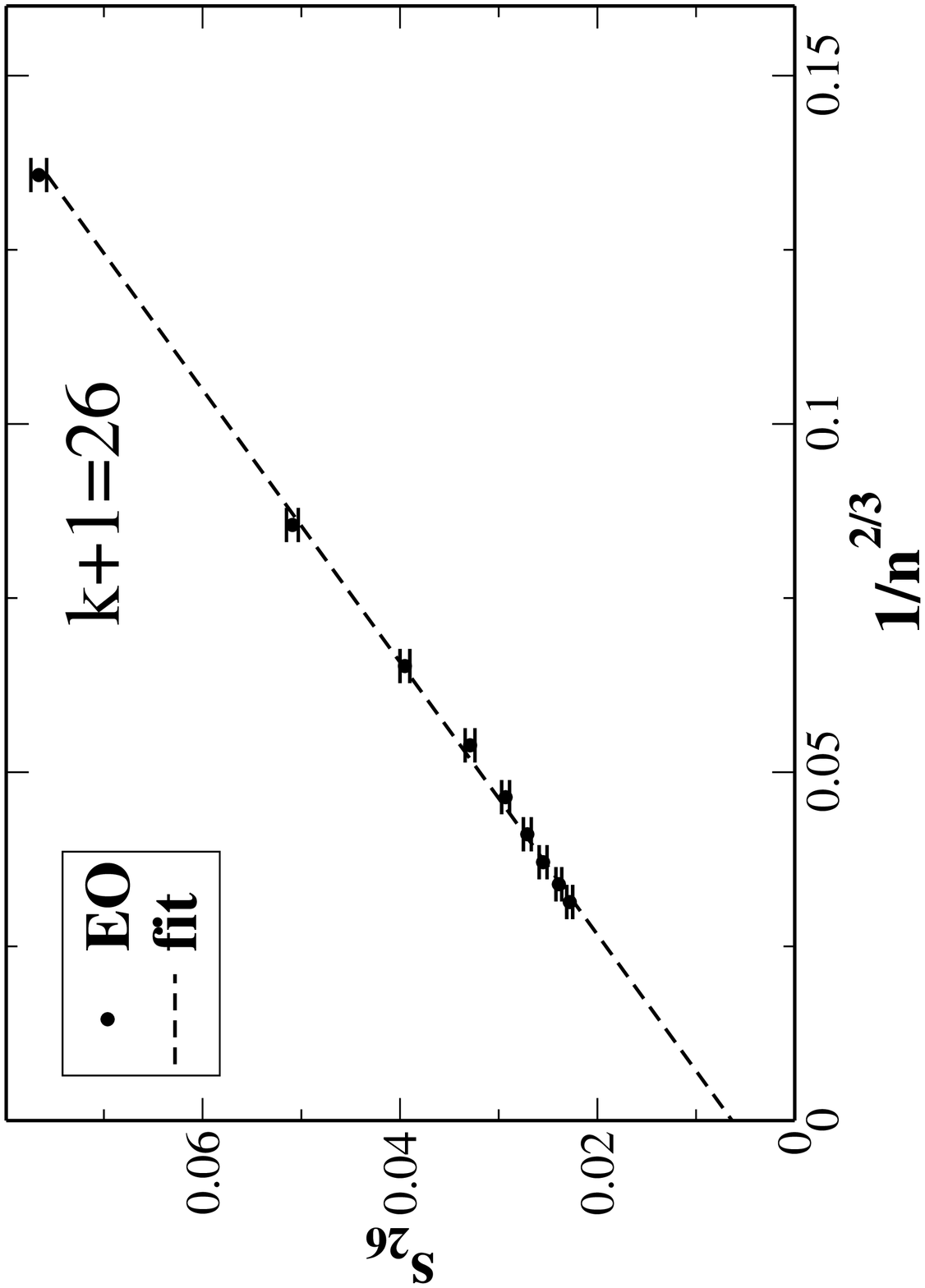} 
\caption{Extrapolation plot for the EO data for the entropy $s_{k+1}(n)$ for some larger, even $k+1$, similar to Figs.~\protect\ref{entroextraplot}.  }
\label{highkentroextraplot}
\end{figure}

Since the range of system sizes $n$ is smaller than for the case of
the energies, it is more difficult to extrapolate our data for
$\left<s_{k+1}(n)\right>$. Again, it is clear that the corrections are
not linear in $1/n$, but instead seems to be scaling close to
$1/n^{2/3}$ for all $k+1$, as for the energies above. Considering the
limitations on $n$, we assume that the corrections are exactly of that
form and extrapolate our data simply with a fit to
\begin{eqnarray}
s_{k+1}(n)\sim s_{k+1}+\frac{A}{n^{2/3}}\quad(n\to\infty),
\label{entroextraeq}
\end{eqnarray}
again, weighting each data point with respect to $n$ and the inverse of
its error. While the systematic and statistical uncertainties of our
data appears to be small, the uncertainty about the scaling
corrections must be considered the most significant limitation on
accuracy in our extrapolation. The data and the extrapolation
fits according to Eq.~(\ref{entroextraeq}) are shown in
Figs.~\ref{entroextraplot} and~\ref{highkentroextraplot}.  The results
for $s_{k+1}$ for $n\to\infty$ are listed in Tab.~\ref{entropy}.

\begin{table}
\caption{Extrapolation results for the entropies per spin for the data plotted in
Figs.~\protect\ref{entroextraplot} and~\protect\ref{highkentroextraplot}.}
\begin{tabular}{rrl|rrl}
\hline\hline
$k+1$ &~~~~&  $s_{k+1}$ & $k+1$ &~~~~&  $s_{k+1}$\\
\hline
3&&	0.0102(10)& 4&&	0.0381(15)\\
5&&	0.0048(10)& 6&&	0.0291(10)\\
7&&	0.0020(10)& 8&&	0.0218(10)\\
9&&	0.0002(15)& 10&&	0.0198(10)\\
15&&     0.0002(15)& 14&&	0.0126(10)\\
                  &&&18&&	0.0095(10)\\
                  &&&22&&	0.0076(10)\\
                  &&&26&&	0.0063(15)\\
\hline\hline
\end{tabular}
\label{entropy}
\end{table}

The data clearly shows a different quantitative behavior between odd
and even values of $k+1$. This difference for the entropies can be
explained in terms of the ``free spins:'' In a highly frustrated
system, even near ground states, many spins are stuck in a situation in
which they violate many of their constraints, no matter how they are
oriented, and changing from one direction to the other may hardly
change the energy of the system. In particular, an even-connected spin
that happens to violate exactly half of its bonds (with $J=\pm1$) can
flip freely without any change to the energy. Odd-connected spins can
only become ``free'' in a connected pair (that happens to violate exactly
half of its {\it external} bonds but satisfies their mutual bond) in
which both simultaneously flip without changing the energy. The latter
situation is naturally far less likely, and thus, purely
even-connected graphs exhibit far more potential for degeneracy at the
ground state than the corresponding odd-connected graphs. Some
preliminary studies for $k+1=3$ and 4 show that in ground state
configurations the fraction of free spins (zero by design for $k+1=3$)
converges to a value just around 5\% for $k+1=4$, while the fraction
of free pairs seems to vanish for large $n$ for both, even and odd
$k+1$. We have not explored the clustering of these states \cite{Hartmann_entro}. We will explore the different behaviors for even and odd $k+1$
in the next Section.

\subsection{Discussion of the Extrapolation Results}
\label{discussion}
\begin{figure}
\vskip 2.6in \includegraphics{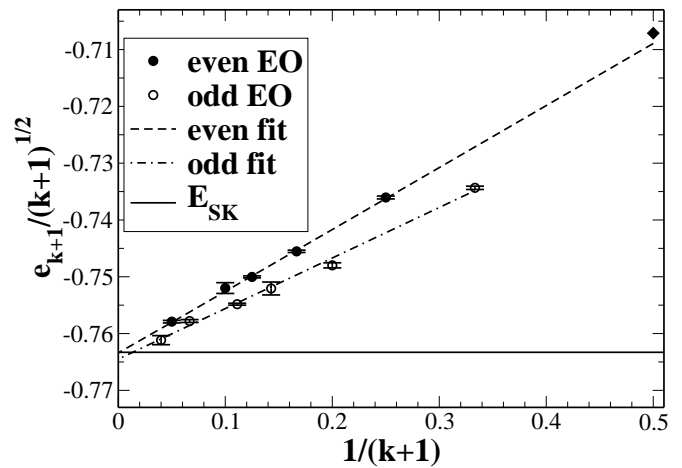}
\caption{Plot of the rescaled extrapolated energies,
$e_{k+1}/\sqrt{k+1}$, as a function of $1/(k+1)$. The data appears to
fall on two separate straight lines for even and for odd $k+1$. The
straight line provides an excellent fit all the way from the exact
result $e_2=-1$ (diamond) to $E_{SK}=-0.7633$ (horizontal line) at
$k+1\to\infty$.  }
\label{SKplot}
\end{figure}

In this section, we want to focus on some of the curious properties
exhibited by the values of the energies and entropies found by
extrapolation in the previous section. We have already noted the
difference between the entropies for even and odd values of $k+1$. In
fact, there are similar differences, although more subtle, for the
energies $e_{k+1}$. These differences become most apparent when we
plot the data asymptotically for large $k+1$, where it is known that
\begin{eqnarray}
\lim_{k+1\to\infty} \frac{e_{k+1}}{\sqrt{k+1}}=E_{SK},
\label{SKeq}
\end{eqnarray}
with $E_{SK}=0.7633$ being the RSB ground state energy of the
Sherrington-Kirkpatrick model \cite{SK,MPV}. In Fig.~\ref{SKplot} we
have plotted $e_{k+1}/\sqrt{k+1}$ as a function of $1/(k+1)$. On this
scale, we notice that the
energies split into a set of even and a set of odd values, each
located apparently on a straight line.  Even though $k+1\leq25$ is quite small, each line separately
extrapolates very close to the exact value for large $k+1$ indeed:
$E_{SK}^{even}\approx-0.763$ and $E_{SK}^{odd}\approx-0.765$. Even
more amazing, the value of $e_2=-1$ [see Eq.~(\ref{twoconnectedeq}) below] for the trivial $k+1=2$ Bethe
lattice  is very close to the linear fit for the even EO
results. Clearly, a function that would interpolate continuously {\it
all} the data will have to be very complicated (oscillatory). But
could it be that its envelope on the even and the odd integers happens
to be simple? Then, in case of the even data \footnote{Although the
odd data may equally well be fitted in this way, the line can not be
determined since only one point on it, $E_{SK}$, is exactly known.},
we could even write down the exact form of the function for ${\cal
E}_{k+1}$ that would fit the data, since we know it also has to pass
$e_2=-1$ and satisfy Eq.~(\ref{SKeq}):
\begin{eqnarray} 
{\cal E}_{k+1}=\sqrt{k+1}E_{SK}-\frac{2E_{SK}+\sqrt{2}}{\sqrt{k+1}}.
\label{eveneq}
\end{eqnarray}

\begin{figure}
\vskip 2.6in  
\includegraphics{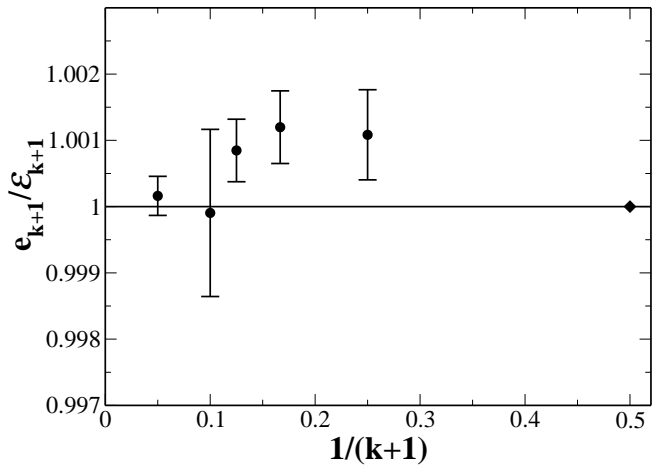}
\caption{Plot of the energies relative to the conjectured function
${\cal E}_{k+1}$ in Eq.~(\protect\ref{eveneq}) as a function of
$1/(k+1)$. All data for even $k+1$ falls within about 0.1\% of ${\cal
E}_{k+1}$ (i.~e. the horizontal line). The point at $k+1=2$ (diamond)
is exact by definition, of course.  }
\label{errorplot}
\end{figure}

\begin{figure}
\vskip 2.6in  
\includegraphics{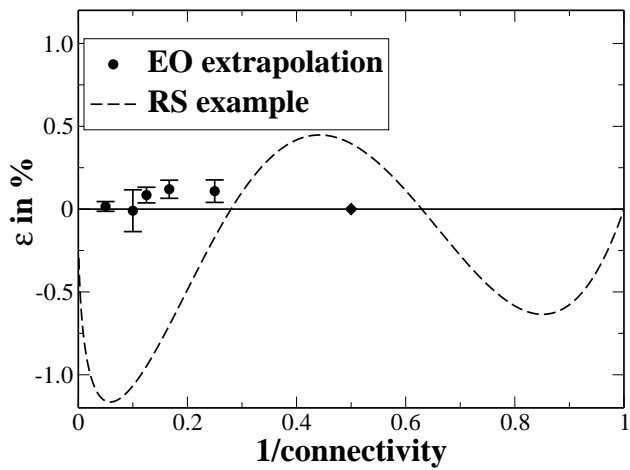}
\caption{Plot of the relative error $\epsilon$ in \% between the
extrapolated data and the function in Eq.~(\protect\ref{eveneq}), as
before in Fig.~\protect\ref{errorplot}, but now superimposed with the
corresponding result obtained for the exactly solvable RS spin glass
on random graphs \protect\cite{KS}. Note the oder-of-magnitude larger
deviations from the reference line for the RS example.  }
\label{kanterplot}
\end{figure}

To test Eq.~(\ref{eveneq}), we plot the data in Fig.~\ref{errorplot}
as $e_{k+1}/{\cal E}_{k+1}$ to study its deviations from the
conjecture. While the extrapolated values do not fall exactly within
their (estimated) error bars on the proposed form, they are indeed
within about 0.1\% of it. To judge how close the data is to the
proposed functional form in Eq.~(\ref{eveneq}), we utilize a closely
related example. The ground-state energy as a function of the
(continuous) average connectivity $\left<c\right>$ is known exactly
for the RS case of ordinary random graphs with fluctuating internal
connectivities, Eq.~(16) in Ref.~\cite{KS}. If one plots that solution 
(which involves exponentials and modified Bessel functions) in the same
way as $e_{k+1}$ in Fig.~\ref{SKplot}, one notes that it, too, could
be approximated surprisingly well with a straight line,
$-\sqrt{2c/\pi}+\left(\sqrt{2/\pi}-\frac{1}{2}\right)/\sqrt{c}$, now
crossing the RS ground state energy $-\sqrt{2/\pi}$ \cite{MPV} for
$\left<c\right>\to\infty$ and reaching the trivial
result of $-1/2$ at the percolation point $\left<c\right>=1$. In
Fig.~\ref{kanterplot} we superimpose the relative error of this
approximation with respect to the exact RS result with the relative error of our data
with respect to the conjecture. It shows that the error of the
conjecture is still almost by an order of magnitude smaller than the
global bound for the RS example, thus putting a significant bound on
any corrections similar in type to Eq.~(16) in Ref.~\cite{KS}.

\begin{figure}
\vskip 2.6in \includegraphics{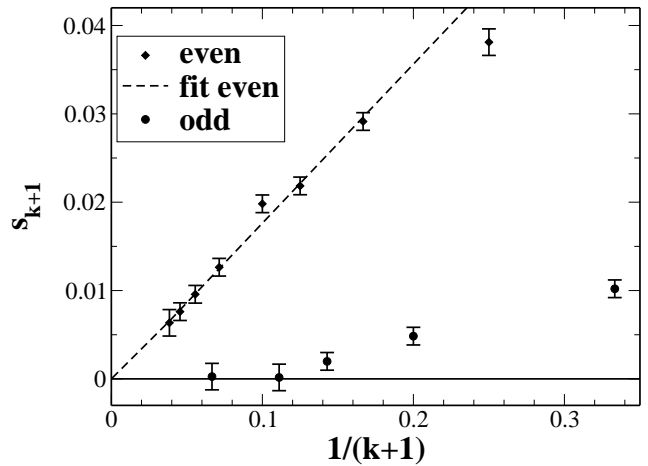}
\caption{Asymptotic plot of the extrapolated entropies from
Tab.~\protect\ref{entropy} as a function of $1/(k+1)$. The data for
even $k+1$ seems to vanish linearly with $1/(k+1)$ (dashed line). The
data for odd $k+1$ drops more precipitously, and can not reasonably be
fitted at this level of accuracy. }
\label{entroscalplot}
\end{figure}

The differences between even and odd connectivities are even more
pronounced in case of the entropies, as we have explained in
Sec.~\ref{numerics}. Thus, although our data for the entropy is not
nearly as accurate as for the energies, it is still instructive to
study it in more detail. In Fig.~\ref{entroscalplot}, we plot the
extrapolated values of the entropies from Tab.~\ref{entropy} to
explore its decrease for large $k+1$. Despite the large error bars, a
significant {\it qualitative} difference between even and odd data
points is visible: The entropy for even values of $k+1$ decays slowly,
apparently linearly with $1/(k+1)$. On the other hand, the entropies
for odd $k+1$ drop much more rapidly, and are already
indistinguishable from zero (within our errors) for $k+1=9$, while it
is clearly non-vanishing for $k+1=3$ [unless our assumption about the
scaling corrections in Eq.~(\ref{entroextraeq}) are incredibly wrong
(see Fig.~\ref{entroextraplot})]. Unfortunately, with only a small,
discrete number of data points available that are significantly above
zero, it is very hard to decide whether the entropy for odd $k+1$
merely decays exponentially, or whether there exists a finite value of
$k+1$ above which all odd entropies become identically zero.

\section{Conclusion}
\label{conclusion}
In this paper we have presented a extensive numerical study of the
ground states of spin glasses on Bethe lattices. The available data
possessed sufficient accuracy to obtain extrapolated values for ground
state energies and entropies at the 0.1\% and the 10\% level,
respectively. In both cases, we significant differences emerged
between the data for odd and even values of $k+1$. Based on the
numerical results, we showed that the extrapolated energies for {\em
all} even values of $2\leq k+1\leq\infty$ was well fitted with a
simple function, Eq.~(\ref{eveneq}). Furthermore, the data suggests
that the entropies are generally non-zero at small $k+1$, but may
vanish above a finite $k+1$ for odd values.
  
Of course, there is plenty of reason to doubt that such a simple
result as Eq.~(\ref{eveneq}), albeit confined to discrete integer
values of $k+1$, could indeed be the solution to a complex RSB
problem. In fact, one argument against the conjecture is a discrepancy
in its prediction for large $k+1$ at next-to-leading order. Several
authors \cite{DG,PT} have studied spin glasses on random graphs beyond
the RS level perturbatively for $k+1=z\to\infty$ to determine the
$1/z$ correction $f_1$ to the free energy (at $T>0$) in
$E_{SK}+f_1/z$. It was recently calculated \cite{PT}, that the
correction for fixed connectivities for $T\to0$ would be about
$f_1=-0.317$, while Eq.~(\ref{eveneq}) would predict about
$0.1124$. It should be noted, though, that the $1/z$ expansion
implicitly assumes a smooth continuation off the integers which may
lead to ambiguities in light of the oscillatory behavior between even
and odd integers we found for the ground states in Fig.~\ref{SKplot}
(similar to the continuation of, say, the function $\cos(2\pi z)/z$
for $z\to\infty$, although this function would not possess a $1/z$
expansion at all).

In any case, future calculations like the one in Ref.~\cite{MP2} but
for even $k+1$ will provide a check on both, our extrapolated data and
the conjecture.

\section*{Acknowledgments}
This work has been supported under a grant from the Emory University
Research Committee. I am greatly indebted to Marc Mezard for his
helpful comments.

\section*{Appendix}
\subsection{The Case $k+1=2$}
Clearly, a Bethe lattice in which each vertex has exactly 2
connections can only consists of a collection of disconnected loop
graphs. We merely need to determine the number of loops and their size
distribution to derive the average ground state energy and
entropy. Each loop has a 50\% chance of being frustrated, thus, the
number of cut bonds is equal to one-half of the number of loops, and
the degeneracy is equal to the length of these loops to the power of
their number.

To analyze the $k+1=2$ case we consider each vertex as a node with two
terminals. Adding lines can create two types of objects: strings and
loops. We consider, after adding $t$ lines, an individual vertex as a
string of length 0, of which there are $l_{0,t}$; in general, we have
$l_{i,t}$ strings of length $i$, each possessing two open
terminals. In particular, before we added any lines:
$l_{i,t=0}=\delta_{i,0}$. A loop of length $i$ is created by addition
of a line to both open terminals of a string of length $i-1$. There
are $p_{i,t}$ loops of length $i$ after adding $t$ lines which can not
evolve further, since they don't possess any more open terminals. We
start with $2n$ open terminals and cover 2 of those with each newly
added line. We can identify two constraints:
\begin{eqnarray}
\sum_{i=0}^\infty \,l_{i,t}=n-t,\qquad \sum_{i=1}^\infty\,i\left(l_{i,t}+p_{i,t}\right)=t.
\label{constrainteq}
\end{eqnarray}

After adding $t$ lines at random, there are $2(n-t)$ terminals left to
accommodate the next line, allowing for ${2\left(n-t\right)\choose2}$
different choices. Accounting for all possible choices, we obtain
\begin{eqnarray}
l_{0,t+1}&=&\left[1-\frac{2}{n-t}+\frac{1}{{2\left(n-t\right)\choose2}}\right]l_{0,t},\nonumber\\
l_{i,t+1}&=&\left[1-\frac{2}{n-t}+\frac{1}{{2\left(n-t\right)\choose2}}\right]l_{i,t}\nonumber\\
&&+\frac{2}{{2\left(n-t\right)\choose2}}\left(\sum_{j=0}^{i-1}l_{j,t}l_{i-1-j,t}-l_{\frac{i-1}{2},t}\vert_{i{\rm
odd}}\right),\nonumber\\
p_{i,t+1}&=&p_{i,t}+\frac{1}{{2\left(n-t\right)\choose2}}l_{i-1,t},
\label{evolutioneq}
\end{eqnarray}
where $i>0$.  It is easy to show that these equations satisfy the
constraints in Eqs.~(\ref{constrainteq}).

We can transform these equations by defining $\theta=t/n$,
$d\theta=1/n$, $y(x,\theta)=\frac{1}{n}\sum_{i=0}^\infty l_{i,t}x^i$,
and $p(x,\theta)=\sum_{i=0}^\infty p_{i,t}x^i$. Considering $n$ large
and $\theta$ continuous, Eqs.~(\ref{evolutioneq}) turn into
\begin{eqnarray}
\frac{dy(x,\theta)}{d\theta}&=&-\frac{2y(x,\theta)}{1-\theta}+\frac{x[y(x,\theta)]^2}{(1-\theta)[1-\theta-1/(2n)]}\nonumber\\
&&+\frac{1}{n}\left[\frac{y(x,\theta)-xy(x^2,\theta)}{2(1-\theta)[1-\theta-1/(2n)]}\right],\nonumber\\
\frac{dp(x,\theta)}{d\theta}&=&\frac{xy(x,\theta)}{(1-\theta)^2},\nonumber\\
&&y(x,0)=1,\qquad p(x,0)=0.
\label{diffeq}
\end{eqnarray}
Luckily, for $n\to\infty$, the equations are easily solved to give
\begin{eqnarray}
y(x,\theta)=\frac{(1-\theta)^2}{1-x\theta},\qquad p(x,\theta)=-\frac{1}{2}\ln(1-x\theta).
\end{eqnarray}

Finally, the total number of loops for the (almost) completed graph, $\theta=1-1/n$, is given by 
\begin{eqnarray}
p(1,1-1/n)=\sum_{i=1}^\infty\,p_{i,n-1}\sim\frac{1}{2}\ln(n).
\end{eqnarray}
On average, half of these loops will be frustrated, i.~e., they will have one of their bonds violated. Since the Hamiltonian in Eq.~(\ref{Heq}) counts the difference between violated and satisfied bonds, or twice the violated bonds minus the number of all bonds, $n(k+1)/2=n$, we get
\begin{eqnarray}
e= \frac{H}{n}\sim-1+\frac{\ln(n)}{2n}.
\label{twoconnectedeq}
\end{eqnarray}

Similarly, we can calculate the degeneracy $\Omega$ of these ground states, roughly,
as the average length of loops, $\left<i\right>=\partial_x\ln p(x,1-1/n)\vert_{x=1}\sim n/\ln(n)$, taken to the power of one-half of their number, $\ln(n)/(4n)$, to give
\begin{eqnarray}
s=\frac{1}{n}\ln\Omega\sim\frac{\ln(n)^2}{4n}.
\end{eqnarray}
Clearly, both the number of violated bonds as well as the entropy vanish in the large-$n$ limit.

\end{document}